\documentclass[twocolumn,english,groupedaddress,amsmath,amssymb,floats,prx]{revtex4-1}
\usepackage[latin9]{inputenc}
\usepackage{color}
\usepackage{mathtools}
\usepackage{amsmath}
\usepackage{graphicx}

\makeatletter

\usepackage{comment}

\@ifundefined{textcolor}{}{%
 \definecolor{BLACK}{gray}{0}
 \definecolor{WHITE}{gray}{1}
 \definecolor{RED}{rgb}{1,0,0}
 \definecolor{GREEN}{rgb}{0,1,0}
 \definecolor{BLUE}{rgb}{0,0,1}
 \definecolor{CYAN}{cmyk}{1,0,0,0}
 \definecolor{MAGENTA}{cmyk}{0,1,0,0}
 \definecolor{YELLOW}{cmyk}{0,0,1,0}
}

\usepackage{epstopdf}
\usepackage{array}
\usepackage{mathrsfs}
\usepackage{dcolumn}
\usepackage{bm}

\newcolumntype{L}[1]{>{\raggedright\let\newline\\\arraybackslash\hspace{0pt}}m{#1}}
\newcolumntype{C}[1]{>{\centering\let\newline\\\arraybackslash\hspace{0pt}}m{#1}}
\newcolumntype{R}[1]{>{\raggedleft\let\newline\\\arraybackslash\hspace{0pt}}m{#1}}

\graphicspath{{../Inkscape/},{../Mathematica/}}

\newcommand{\mbf}[1]{\mathbf{#1}}

\usepackage{tikz}
\usetikzlibrary{calc,intersections}
\usetikzlibrary{shadings}
\usepgflibrary{shadings}

\usepackage{babel}

\usepackage{babel}

\makeatother

\usepackage{babel}
\begin{document}

\title{Intertwined spin-orbital coupled orders in the iron-based superconductors}

\author{Morten H. Christensen}
\email{mchrist@umn.edu}

\author{Jian Kang}
\altaffiliation{Present Address: National High Magnetic Field Laboratory, Florida State University, Tallahassee, Florida 32304, USA}

\author{Rafael M. Fernandes}

\affiliation{School of Physics and Astronomy, University of Minnesota, Minneapolis,
Minnesota 55455, USA}

\date{\today}
\begin{abstract}
The underdoped phase diagram of the iron-based superconductors exemplifies
the complexity common to many correlated materials. Indeed, multiple
ordered states that break different symmetries but display comparable
transition temperatures are present. Here, we argue that such a complexity
can be understood within a simple unifying framework. This framework,
built to respect the symmetries of the non-symmorphic space group
of the FeAs/Se layer, consists of primary magnetically-ordered states
and their vestigial phases that intertwine spin and orbital degrees
of freedom. All vestigial phases have Ising-like and zero wave-vector
order parameters, described in terms of composite spin order and exotic
orbital-order patterns such as spin-orbital loop-currents, staggered
atomic spin-orbit coupling, and emergent Rashba- and Dresselhaus-type
spin-orbit interactions. Moreover, they host unusual phenomena, such
as the electro-nematic effect, by which electric fields acts as transverse
fields to the nematic order parameter, and the ferro-Néel effect,
by which a uniform magnetic field induces Néel order. We discuss the
experimental implications of our findings to iron-based superconductors
and possible extensions to other correlated compounds with similar
space groups. 
\end{abstract}
\maketitle

\section{Introduction}

The complexity of the phase diagrams of correlated systems often challenges
the notion of a unifying, simple framework to describe these fascinating
systems. One example is that of the underdoped hole-doped cuprates:
besides the mysterious pseudogap phenomenon, they display a plethora
of ordered phases \textendash{} incommensurate magnetism, charge order,
nematicity, time-reversal symmetry-breaking order, and inversion symmetry-breaking
order~(for a recent review, see Ref.~\onlinecite{keimer15}). As pointed out
in Ref.~\onlinecite{tranquada15}, it is difficult to explain this richness
solely in terms of independent, competing electronic orders. This
led to proposals of a more fundamental type of ordered state, such
as a pair-density wave~\cite{himeda02,berg07,agterberg08,lee14},
which simultaneously breaks several of the symmetries above-mentioned.
In this scenario, the various broken-symmetry phases can be interpreted
as vestigial orders that break only a subset of the symmetries of
this ``mother'' state. To contrast with the standard case of competing
orders and fine-tuned multicritical points, they have been dubbed
intertwined orders~\cite{tranquada15,fernandes18}.

\begin{figure*}
\centering \includegraphics[width=1\textwidth]{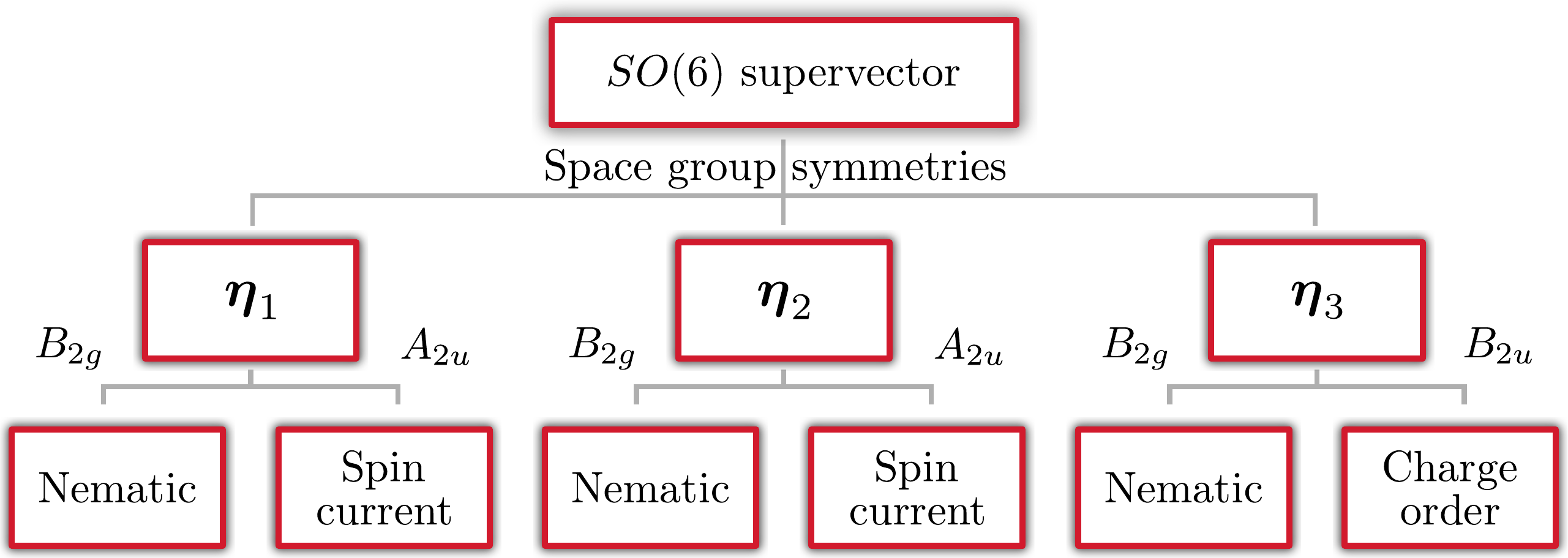}
\caption{\label{fig:intertwined_schematic} Schematic representation of the
cascade of symmetry breakings leading to the emergence of intertwined
orders in the iron-based superconductors. The starting point is the
magnetic $SO(6)$ super-vector, presented in Eq. (\ref{supervector}).
Space-group symmetries only allow certain combinations of the super-vector
components to order, giving rise to the primary magnetically ordered
states $\boldsymbol{\eta}_{i}$. These, in turn, support different
types of vestigial spin-orbital coupled phases, characterized by Ising-like,
zero wave-vector order parameters that transform as the $B_{2g}$,
$A_{2u}$, and $B_{2u}$ irreducible representations of the tetragonal
point group.}
\end{figure*}

The phase diagram of underdoped iron-based superconductors seems relatively
simpler compared to their copper-based counterparts~\cite{paglione10,johnston10,mazin11,chubukov12}.
Initial observations led to the concept of a typical phase diagram
displaying two ordered normal states (besides superconductivity),
namely, stripe magnetism and nematicity~\cite{nandi10} (a notable
exception is FeSe~\cite{bohmer17,coldea18}, which we will discuss
more later). One possible scenario is that these two ordered states
have different microscopic origins~\cite{ku09,kruger09,khalyavin14}.
However, early on, it was argued that the nematic state can be understood
as a vestigial phase of the stripe magnetic state, as it breaks a
subset of the symmetries broken by the latter \textendash{} specifically,
tetragonal symmetry \cite{fang08,xu08,batista11,fernandes12}. In
this scenario, stripe magnetism would be the ``mother'' phase of
underdoped iron pnictides~\cite{fernandes14a}. Nematic order is
manifested in the electronic spectrum as orbital order involving the
$3d_{xz}$, $3d_{yz}$, and $3d_{xy}$ Fe orbitals that form the low-energy
electronic states. The vestigial nematic state in the iron pnictides
is therefore a prime example of an intertwined spin-orbital coupled
phase \cite{Dagotto13,fanfarillo15}, characterized by a composite
spin order parameter and a simple (i.e. non-composite) orbital order
parameter.

Analogously to the historical evolution of the ``typical'' phase
diagram of the cuprates, more detailed experiments in the iron-based
materials have recently unveiled a much more intricate underdoped
phase diagram. In particular, besides stripe magnetism and nematicity,
other types of magnetic order were observed to proliferate as optimal
doping is approached~\cite{wang16}. These are magnetic configurations
that do not break tetragonal symmetry~\cite{kim10,hassinger12,avci14,bohmer15,waser15,allred16,hassinger16,taddei17,taddei16,meier18,bohmer18},
and have thus been dubbed $C_{4}$ magnetic phases. This phenomenon
is observed quite broadly in hole-doped pnictides, and also in certain
materials under pressure~\cite{hassinger12,hassinger16,bohmer18,kasanov18}.
These observations called for revisiting the notion that stripe magnetism
may be the ``mother'' phase of underdoped iron superconductors.
A compelling scenario, put forward by several groups~\cite{lorenzana08,eremin08,brydon11,fernandes12,cvetkovic13,kang14,wang15,gastiasoro15,christensen15,christensen17},
is that the $C_{4}$ phases are double-$\mbf{Q}$ magnetic states,
while the stripe phase is a single-$\mbf{Q}$ magnetic state. Here,
$\mathbf{Q}$ refers to two ordering vectors related by a $90^{\circ}$
rotation; in the square unit cell containing one Fe atom only, $\mathbf{Q}_{1}=\left(\pi,0\right)$
and $\mathbf{Q}_{2}=\left(0,\pi\right)$. In this scenario, all magnetically
ordered phases are described in terms of two magnetic vector order
parameters: 
\begin{equation}
\mathbf{M}_{1}=\left(\begin{array}{c}
M_{1,x}\\
M_{1,y}\\
M_{1,z}
\end{array}\right)\,,\,\mathbf{M}_{2}=\left(\begin{array}{c}
M_{2,x}\\
M_{2,y}\\
M_{2,z}
\end{array}\right)\,.\label{M}
\end{equation}
Formally, $\mathbf{M}_{a}$ is the staggered Fe magnetization with
momentum $\mathbf{Q}_{a}$. One can then consider the ``mother''
order parameter as a hypothetical six-dimensional super-vector $\boldsymbol{\mathcal{M}}=\left(\mathbf{M}_{1},\,\mathbf{M}_{2}\right)^{T}$.
The symmetries of the square lattice only allow certain combinations
of the six components of $\boldsymbol{\mathcal{M}}$ to condense,
namely, one single-$\mbf{Q}$ state (the stripe magnetic order) and
two double-$\mbf{Q}$ states (called spin-vortex crystal~\cite{meier18}
and charge-spin density-wave~\cite{allred16}).

The single-$\mbf{Q}$ phase breaks tetragonal symmetry and is therefore
associated with a vestigial nematic phase. An important question is
what types of vestigial orders can be associated with the double-$\mbf{Q}$
magnetic phases. This question was partially addressed in Ref.~\onlinecite{fernandes16},
which proposed unusual vestigial phases that break translational and/or
mirror symmetries, while preserving time-reversal symmetry. There
is an important issue left unaddressed, however: what are the orbital-order
patterns, if any, related to these vestigial phases? As discussed
above, one of the hallmarks of nematicity is precisely its accompanying
Fe-orbital order.

In this paper, we use group theory to determine the orbital ordering
configurations of these vestigial phases, revealing a rich landscape
of intertwined spin-orbital coupled phases. More than just an interesting
extension of previous results, our work provides a powerful new framework
to describe and predict possible ordered phases of underdoped iron-based
superconductors, as well as their responses to external electromagnetic
and strain fields. The key point is that the description of the magnetic
degrees of freedom in terms of the order parameters $\mathbf{M}_{1}$
and $\mathbf{M}_{2}$ is inevitably incompatible with a description
of the orbital degrees of freedom, due to the crystal symmetry of
the FeAs (or FeSe) layer. More specifically, while the magnetic properties
alone can be reasonably described in terms of a simplified unit cell
containing only one Fe atom, the five $3d$ Fe-orbitals cannot.

\begin{figure}
\centering \includegraphics[width=0.9\columnwidth]{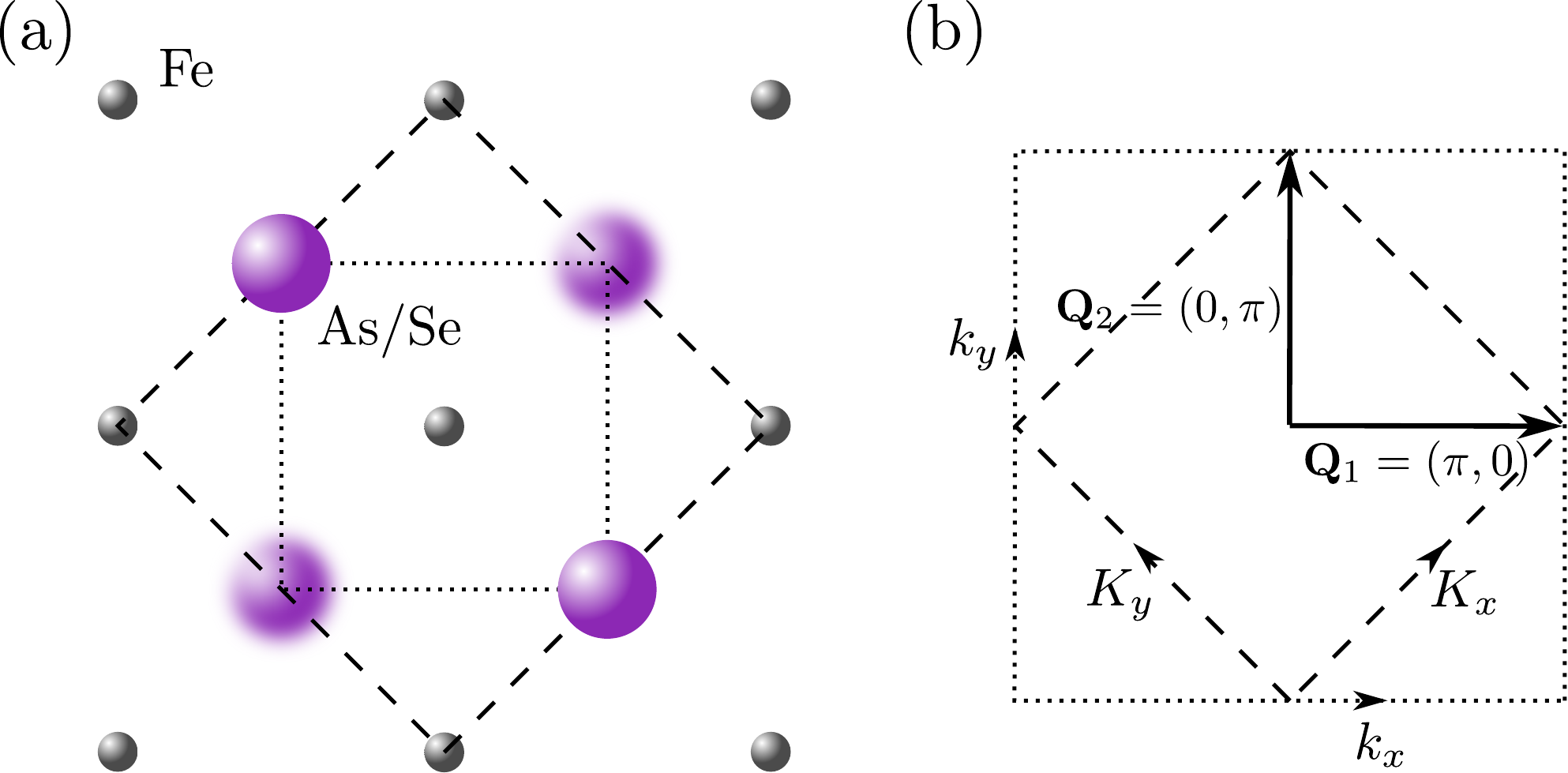} \caption{\label{fig:unit_cell} (a) Illustration of the 1-Fe (dotted square)
and 2-Fe (dashed square) unit cells. To correctly capture the puckering
of the As/Se atoms, two Fe atoms and two As/Se atoms are required in the unit cell. Here
sharp purple spheres denote As/Se atoms above the Fe-layer while blurred
purple spheres denote As/Se atoms below the Fe-layer. (b) Crystallographic
2-Fe Brillouin zone (dashed square) and idealized 1-Fe Brillouin zone
(dotted square).}
\end{figure}

The resolution to this conundrum is to note that the presence of a
sizable spin-orbit coupling, as observed experimentally in most iron-based
superconductors~\cite{borisenko16}, ties the crystal symmetry properties
of the spin degrees of freedom to those of the orbitals. Group-theoretical
analysis of the space group of single-layered FeAs/Se superconductors
($P4/nmm$) then implies that the six components of $\boldsymbol{\mathcal{M}}$
should not be grouped in terms of two three-dimensional vectors with
different wave-vectors, Eq. (\ref{M}), but to three two-dimensional
vectors with the same wave-vector~\cite{cvetkovic13,christensen15}:
\begin{equation}
\boldsymbol{\eta}_{1}=\left(\begin{array}{c}
M_{1,x}\\
M_{2,y}
\end{array}\right)\,,\,\boldsymbol{\eta}_{2}=\left(\begin{array}{c}
M_{1,y}\\
M_{2,x}
\end{array}\right)\,,\,\boldsymbol{\eta}_{3}=\left(\begin{array}{c}
M_{1,z}\\
M_{2,z}
\end{array}\right)\,.\label{eq:eta}
\end{equation}
In particular, while $\mathbf{M}_{a}$ lives in the artificial one-Fe
square lattice unit cell ($a=1,\,2$), $\boldsymbol{\eta}_{\alpha}$
lives in the actual crystallographic unit cell of a single layer ($\alpha=1,\,2,\,3$),
containing two Fe atoms and two As/Se atoms. Because both wave-vectors
$\mathbf{Q}_{1}=\left(\pi,0\right)$ and $\mathbf{Q}_{2}=\left(0,\pi\right)$
of the 1-Fe Brillouin zone map onto the same wave-vector $\mathbf{Q}_{M}=\left(\pi,\pi\right)$
of the 2-Fe Brillouin zone, all $\boldsymbol{\eta}_{\alpha}$ have
the same wave-vector, as seen in Fig.~\ref{fig:unit_cell}. Formally,
they correspond to three two-dimensional irreducible representations
of the FeAs single-layer space-group. 

The important point is that any low-energy field-theory for the magnetic
degrees of freedom should be formulated in terms of $\boldsymbol{\eta}_{\alpha}$,
rather than $\mathbf{M}_{a}$, in contrast to what has been done extensively
in the previous literature. Applying this formalism to classify the
possible magnetically-driven vestigial orders, we find three distinct
$\mathbf{Q}=0$ orbitally-ordered states, characterized by Ising-like
order parameters that are composite combinations of $\boldsymbol{\eta}_{\alpha}$
but simple (i.e. non-composite) combinations of orbital operators.
They are: 
\begin{enumerate}
\item The well-studied $\mathbf{Q}=0$ nematic phase, whose Ising order
parameter transforms as the $B_{2g}$ irreducible representation of
the tetragonal group. This is the vestigial phase of the stripe magnetic
state. The corresponding orbital order configuration is a mixture
of onsite and bond $d_{xz}/d_{yz}$ ferro-orbital order, as well as
$d_{xy}$ bond order. Structurally, this order couples to an acoustic
phonon mode, and is thus accompanied by an orthorhombic lattice distortion.
Its conjugate field is shear strain along the Fe-Fe bonds.
\item A $\mathbf{Q}=0$ ordered phase whose Ising order parameter transforms
as the $B_{2u}$ irreducible representation of the tetragonal group.
This is the vestigial phase of the charge-spin density-wave state.
The corresponding orbital order configuration is a combination of
intra-unit-cell staggered charge order on the $d_{xy}$ orbitals and
a intra-unit-cell ``staggered spin-orbit coupling'' ordering involving
the $d_{xz}/d_{yz}$ orbitals. This ordered state does not couple
to any new atomic displacements, but it does remove the spin-degeneracy
of the bands, giving rise to spontaneously generated Dresselhaus-type
and Rashba-type spin-orbit terms. For simplicity, we refer to this
phase as the charge-ordered state. Its conjugate field is a combination
of an electric field perpendicular to the FeAs-layer and shear strain
along the Fe-As bonds. 
\item A $\mathbf{Q}=0$ ordered phase whose Ising order parameter transforms
as the $A_{2u}$ irreducible representation of the tetragonal group.
This is the vestigial phase of the spin-vortex crystal state. The
corresponding orbital order configuration is a mixture of staggered
spin-current patterns involving the $d_{xy}$ orbitals and a mirror-symmetry
breaking ordering involving the $d_{xz}/d_{yz}$ orbitals. This type
of order couples to an optical phonon mode, whose softening however
does not lead to any new atomic displacements. Similarly to the $B_{2u}$
Ising order, it also removes the spin-degeneracy of the bands, triggering
the appearance of Dresselhaus-type and Rashba-type spin-orbit terms.
For simplicity, we refer to this phase as the spin-current state.
Its conjugate field is an electric field perpendicular to the FeAs-layer. 
\end{enumerate}
A summary of the framework derived here, including the primary magnetic
states and their vestigial phases is shown in Fig. \ref{fig:intertwined_schematic}.
Because these three vestigial phases are not independent, but connected
by transformations in the two-dimensional internal spaces of the $\boldsymbol{\eta}_{\alpha}$
order parameters, they can act as ``transverse fields'' to each
other~\cite{maharaj17}. This opens the possibility of using, for
instance, an electric field to induce the spin-current vestigial order,
which can then be employed to tune the nematic phase transition. This
rather unusual effect, which we dub \emph{electro-nematic effect},
is intimately connected to the intertwined character of phases that
break completely different symmetries.

The same formalism can be used to investigate how the vestigial phases
respond to external fields, such as a magnetic field $\mathbf{H}$.
In the nematic phase, the field simply induces a finite magnetization,
as expected for a standard paramagnet. However, in the charge-ordered
and spin-current states, the field $\mathbf{H}$ also induces a finite
N{é}el magnetization $\mathbf{N}$, despite the absence of any long-range
magnetic order. In the charge-ordered state, $\mathbf{N}\parallel\mathbf{H}$,
resulting in a ferrimagnetic spin configuration, whereas in the spin-current
state we find that an in-plane magnetic field induces a canted spin
configuration for $\mbf{H}\parallel\hat{\mathbf{x}}$ or $\mbf{H}\parallel\mathrm{\hat{\mathbf{y}}}$,
and a ferrimagnetic configuration if $\mbf{H}$ is applied along the
diagonal direction of the 1-Fe unit cell. This effect, which we dub
\emph{ferro-Néel effect}, shows the non-trivial character of these
vestigial paramagnetic phases, and can be used to identify these exotic
spin-orbit coupled vestigial states.

This paper is organized as follows: In Sec.~\ref{sec:vestigial_phases}
we discuss how vestigial phases arise in the iron-based systems. We
argue that the fundamental degrees of freedom are given by Eq.~\eqref{eq:eta},
and write down the field theory invariant under the space-group symmetries
of the FeAs/Se layer. We demonstrate that vestigial phases can arise
in the paramagnetic state, and are characterized by composite spin
order. In Sec.~\ref{sec:el_impact}, we derive the electronic orders
arising as consequences of the vestigial phases, and demonstrate how
this leads to intertwined spin-orbital coupled orders. The electronic
orders induced in this manner include the well-known ferro-orbital
order arising in the nematic case, but also a more exotic spin-current
order. We show how a number of these induced electronic orders lead
to the lifting of the spin-degeneracy of the bands. The vestigial
phases exhibit unique features when subjected to external electromagnetic
fields. These are discussed in Sec.~\ref{sec:ind_orders}. In Sec.~\ref{sec:diff_11_122}
we discuss how our approach changes depending on the stacking of the
FeAs/Se layers, contrasting 1111, 111, 11 compounds and 122 compounds.
Finally, conclusions are presented in Sec.~\ref{sec:conclusions}.
Additionally, we include two appendices containing technical details
of the Hamiltonian used (Appendix~\ref{app:kp_model}) and the derivation
of the fermionic bilinears (Appendix~\ref{app:fermionic_terms}).

\section{Composite spin order in the vestigial phases}

\label{sec:vestigial_phases}

\subsection{Magnetically ordered states}

Recent experiments unveiled an unexpected complexity of the phase
diagram of underdoped iron-based superconductors, which rivals that
of underdoped cuprates. Our goal is to formulate a unifying framework
in which the observed magnetic and non-magnetic orders arise from
the same basic interactions, thus providing an organizing principle
to describe and predict the behavior of underdoped iron-based superconductors
without resorting to fine-tuned multi-critical points. The main idea
is that all these phases can be described as condensations of simple
or composite combinations of the components of the six-dimensional
super-vector: 
\begin{equation}
\boldsymbol{\mathcal{M}}=\left(\begin{array}{cccccc}
M_{1,x} & M_{1,y} & M_{1,z} & M_{2,x} & M_{2,y} & M_{2,z}\end{array}\right)\,.\label{supervector}
\end{equation}
Here, $M_{a,\mu}$ are the components of the staggered magnetizatons
$\mathbf{M}_{1}$ and $\mathbf{M}_{2}$ given in Eq.~\eqref{M}.
In this framework, the primary degrees of freedom are the magnetic
ones. This does not mean that orbital degrees of freedom are irrelevant
\textendash{} much to the contrary, they are essential to correctly
describe the vestigial phases, as shown in Ref. \onlinecite{fanfarillo15}.
Empirically, it is well established that most iron superconductors
display sharp magnetic fluctuations around the wave-vectors $\mathbf{Q}_{1}=(\pi,0)$
and $\mathbf{Q}_{2}=(0,\pi)$~\cite{dai15}. Microscopically, one
expects on general grounds that repulsive electronic interactions
are responsible for the enhancement of these fluctuations~\cite{eremin08}.
In this paper, we will not discuss the microscopic origin of this
magnetism. Instead, we simply note that both weak-coupling~\cite{lorenzana08,eremin08,fernandes12,gastiasoro15,glasbrenner15,christensen17}
and strong-coupling~\cite{si08,medici11,medici14,DHLee15} approaches
give magnetic fluctuations peaked at these ordering vectors \cite{Calderon16}.
\begin{figure}
\includegraphics[width=1\columnwidth]{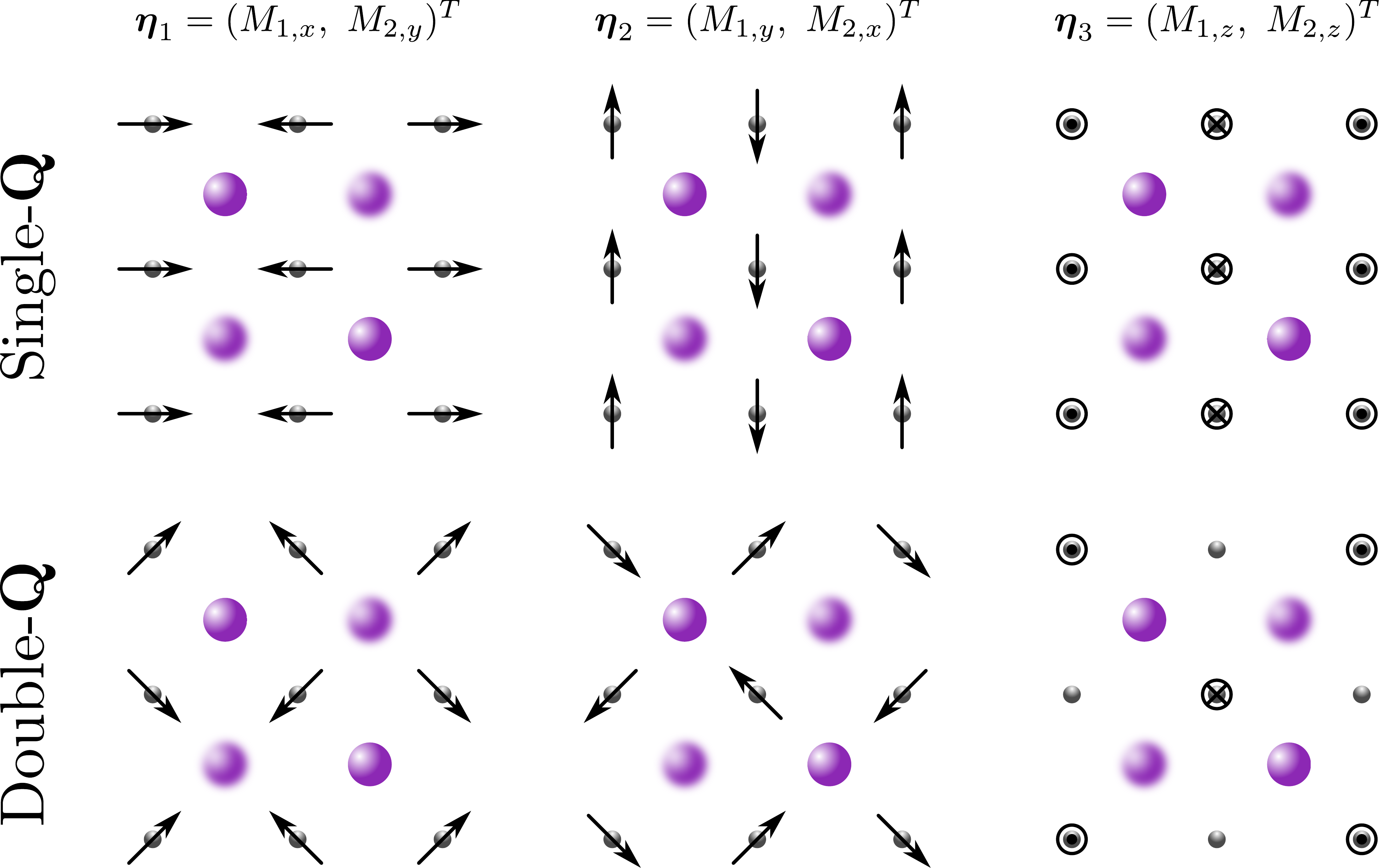}
\caption{\label{fig:mag_order} Illustration of the possible magnetic orders
for each order parameter $\boldsymbol{\eta}_{i}$, with transforms
as a two-dimensional irreducible representation of the space group.
The double-$\mbf{Q}$ structures for $\boldsymbol{\eta}_{1}$ and
$\boldsymbol{\eta}_{2}$ are the in-plane hedgehog- and loop-SVC phases,
respectively. For $\boldsymbol{\eta}_{3}$, the double-$\mbf{Q}$
phase is denoted the charge-spin density-wave phase. All single-\textbf{Q
}phases are stripe magnetically ordered states.}
\end{figure}

In the absence of a lattice, the elements of $\boldsymbol{\mathcal{M}}$
would transform as an irreducible representation (irrep) of the $SO(6)$
group. There is a huge degeneracy of magnetic ground states that give
the same amplitude of the super-vector $\boldsymbol{\mathcal{M}}$.
The vast majority of these states are not realized in the iron-based
superconductors. This reflects the important fact that this enlarged
$SO(6)$ symmetry is merely a theoretical construct, as the existence
of the lattice immediately constrains the possible combinations of
the elements of $\boldsymbol{\mathcal{M}}$ to those that obey the
crystal lattice symmetries. We stress that the magnetic order parameters are not SO(6) invariant. Below we use group theory to show how the crystal lattice restricts the symmetry of the magnetic order parameters.

In the simple approximation where the As/Se atoms are neglected, and
the crystal is described as a single-Fe square lattice, the six components
of $\boldsymbol{\mathcal{M}}$ do not transform according to an irreducible
representation of the $SO(6)$ group, but instead according to an
irrep of the $C_{4v}^{'''}\otimes SO(3)$ group, as discussed in Ref.
\onlinecite{fernandes18}. Here, $C_{4v}^{'''}$ is the extended point group
corresponding to the point group $C_{4v}$ supplemented by three translations
along $\hat{\mathbf{x}}$, $\hat{\mathbf{y}}$, and $\hat{\mathbf{x}}+\hat{\mathbf{y}}$.
The translations are necessary because the order parameters $M_{a,\mu}$
break the translational symmetry of the lattice. Note that, for our
purposes, we neglected the inversion symmetry and considered $C_{4v}$
instead of $D_{4h}$. As discussed in Ref.~\onlinecite{fernandes18},
the six elements of $\boldsymbol{\mathcal{M}}$ transform as the product
of irreps $E_{5}\otimes\Gamma^{S=1}$, where $E_{5}$ is a two-dimensional
irreducible representation of $C_{4v}^{'''}$ and $\Gamma^{S=1}$
is the standard 3-dimensional irreducible representation of $SO(3)$.
This warrants writing $\boldsymbol{\mathcal{M}}=\left(\mathbf{M}_{1},\,\mathbf{M}_{2}\right)^{T}$
and identifying the appropriate order parameters as the three-dimensional
vectors $\mathbf{M}_{1}$ and $\mathbf{M}_{2}$, instead of $\boldsymbol{\mathcal{M}}$.
The resulting free energy, which can be constructed by imposing that
its elements transform trivially under the group symmetry operations,
becomes: 
\begin{align}
F\left[\mathbf{M}_{a}\right] & =\frac{a}{2}\left(M_{1}^{2}+M_{2}^{2}\right)+\frac{u}{4}\left(M_{1}^{2}+M_{2}^{2}\right)^{2}\nonumber \\
 & -\frac{g}{4}\left(M_{1}^{2}-M_{2}^{2}\right)^{2}+w\left(\mathbf{M}_{1}\cdot\mathbf{M}_{2}\right)^{2}\,.\label{F_M}
\end{align}
Clearly, while the first two terms only depend on the amplitude of
the super-vector $\boldsymbol{\mathcal{M}}$, the last two terms explicitly
break $SO(6)$ symmetry. This free-energy expansion, and the field
theory resulting from it, have been widely studied in several papers
in the context of the pnictides~\cite{lorenzana08,brydon11,fernandes12,kang14,wang15,fanfarillo15,christensen16,christensen17}.
We just quote here the result that there are only three possible ground
states, which are determined by the quartic coefficients $g$ and
$w$: the single-\textbf{Q }stripe magnetic state, where $\mathbf{M}_{1}\neq0$
and $\mathbf{M}_{2}=0$ (or vice-versa); the non-collinear double-\textbf{Q
}spin-vortex crystal state, where $M_{1}=M_{2}$ and $\mathbf{M}_{1}\perp\mathbf{M}_{2}$;
and the collinear double-\textbf{Q }charge-spin density-wave, where
$M_{1}=M_{2}$ and $\mathbf{M}_{1}\parallel\mathbf{M}_{2}$. All these
states have been observed in the iron-based superconductors~\cite{huang08,cruz08,dong08,allred16,meier18}.

Although the magnetic properties of these systems are decently described
within the approximation of a single-Fe square lattice, the symmetries
of the actual FeAs/Se layer are different than those of the simple
square lattice. Most notably, there is a glide plane symmetry originating
from the puckering of the As/Se atoms above and below the Fe plane.
The immediate consequence of including the As/Se atoms is that the
unit cell doubles and rotates by $45^{\circ}$ (see Fig.~\ref{fig:unit_cell}).
Note that this unit cell is invariant under a four-fold rotation only
if it is accompanied by a mirror reflection. There are also important
consequences for the electronic structure, as the different Fe $3d$
orbitals do not necessarily transform the same way under the glide
plane symmetry operation \cite{ku11,valenti14}. Crucially, a sizable
atomic spin-orbit coupling $\lambda\mathbf{S}\cdot\mathbf{L}$ is
observed in these materials ($\lambda\sim20$ meV~\cite{borisenko16}).
As a result, the spin-space $SO(3)$ symmetry is broken, and the six
magnetic components no longer transform according to an irrep of the
$C_{4v}^{'''}\otimes SO(3)$ group. Instead, the relevant group is
the space-group of a single FeAs/Se layer, which is the non-symmorphic
$P4/nmm$ group (group number 129)~\cite{cvetkovic13,khalyavin14}.

In this case, the six elements of $\boldsymbol{\mathcal{M}}$ cannot
be organized in the two three-dimensional vectors $\mathbf{M}_{1}$
and $\mathbf{M}_{2}$. Due to the doubling and $45^{\circ}$ rotation
of the Fe-square unit cell, the two wave-vectors $\mathbf{Q}_{1}=\left(\pi,0\right)$
and $\mathbf{Q}_{2}=\left(0,\pi\right)$ are folded onto the same
wave-vector $\mathbf{Q}_{M}=\left(\pi,\pi\right)$ (see Fig.~\ref{fig:unit_cell}).
As a result, the six elements of $\boldsymbol{\mathcal{M}}$ must
transform according to different irreps of the space group $P4/nmm$
at the $M\equiv\left(\pi,\pi\right)$ point of the crystallographic
Brillouin zone. It was shown in Ref.~\onlinecite{cvetkovic13} that
all irreps of $P4/nmm$ at $M$ are two-dimensional. As a result,
the six elements of $\boldsymbol{\mathcal{M}}$ should be organized
in terms of three different two-dimensional irreps, denoted by $E_{M1}$,
$E_{M2}$, and $E_{M3}$ (using the terminology of Ref.~\onlinecite{cvetkovic13}).
The corresponding two-dimensional order parameters are, respectively,
the $\boldsymbol{\eta}_{1}$, $\boldsymbol{\eta}_{2}$, and $\boldsymbol{\eta}_{3}$
vectors given by Eq.~\eqref{eq:eta}. Their physical meaning can
be understood in a straightforward way: in the presence of spin-orbit
coupling, $M_{1,x}$, $M_{1,y}$, and $M_{1,z}$ become independent.
Each of these order parameters have a partner related by a simultaneous
$90^{\circ}$ rotation around $\hat{\mbf{z}}$ in both real and spin
spaces, corresponding to $M_{2,y}$, $M_{2,x}$, and $M_{2,z}$.

The field theory for the magnetic degrees of freedom should thus be
expressed in terms of the two-component vectors $\boldsymbol{\eta}_{\alpha}$,
rather than the three-component vectors $\mathbf{M}_{a}$. Imposing
that the free-energy expansion transforms trivially under the operations
of the space group $P4/nmm$, we find: 
\begin{align}
F\left[\boldsymbol{\eta}_{\alpha}\right] & =\sum_{\alpha}\frac{a_{\alpha}}{2}\left(\boldsymbol{\eta}_{\alpha}\tau^{0}\boldsymbol{\eta}_{\alpha}\right)+\sum_{\alpha,\beta}\frac{u_{\alpha\beta}}{4}\left(\boldsymbol{\eta}_{\alpha}\tau^{0}\boldsymbol{\eta}_{\beta}\right)^{2}\nonumber \\
 & -\sum_{\alpha,\beta}\frac{g_{\alpha\beta}}{4}\left(\boldsymbol{\eta}_{\alpha}\tau^{z}\boldsymbol{\eta}_{\beta}\right)^{2}+\sum_{\alpha\neq\beta}\frac{w_{\alpha\beta}}{4}\left(\boldsymbol{\eta}_{\alpha}\tau^{x}\boldsymbol{\eta}_{\beta}\right)^{2}\,.\label{F_eta}
\end{align}
Here, $\alpha,\beta=1,\,2,\,3$ and $\tau^{j}$ are Pauli matrices
that live in the two-dimensional ``internal'' space of $\boldsymbol{\eta}_{\alpha}$.
Symmetry constrains a number of the coefficients, e.g. $u_{13}+g_{13}=-w_{13}$
and $u_{23}+g_{23}=-w_{23}$. Note that the quadratic term was previously
obtained in Refs.~\onlinecite{christensen15,scherer18}. Moreover,
the $w_{\alpha\beta}$-coefficients play no role at the magnetic transition~\cite{christensen18a,christensen18b}.
We emphasize that several of the magnetic properties derived from
Eq.~\eqref{F_M} still hold in this alternative formulation. But
as we show below, the field theory that respects the actual crystallographic
symmetries of the FeAs/Se layers provides unique insights into the
rich interplay between spin and orbital degrees of freedom.

The six different ground states of Eq.~\eqref{F_eta} are obtained
by direct minimization (see also Ref.~\onlinecite{christensen15}),
and are shown in Fig.~\ref{fig:mag_order}. For each of the three
irreps, represented by $\boldsymbol{\eta}_{\alpha}$, there are two
possible magnetic ground states, corresponding to single-\textbf{Q
}configurations (i.e. condensation of one of the components of $\boldsymbol{\eta}_{\alpha}$)
or double-\textbf{Q }configurations (i.e. simultaneous condensation
of both components of $\boldsymbol{\eta}_{\alpha}$). Specifically,
the ground states associated with $\boldsymbol{\eta}_{1}$ are the
single-\textbf{Q} stripe magnetic phase with moments pointing parallel
to the wave-vectors $\mathbf{Q}_{1}=\left(\pi,0\right)$ and $\mathbf{Q}_{2}=\left(0,\pi\right)$,
and the double-\textbf{Q }hedgehog spin-vortex crystal phase~\cite{meier18}.
Similarly, the ground states associated with $\boldsymbol{\eta}_{2}$
are the single-\textbf{Q} stripe magnetic phase with moments pointing
in-plane, but perpendicular to the wave-vectors $\mathbf{Q}_{1}=\left(\pi,0\right)$
and $\mathbf{Q}_{2}=\left(0,\pi\right)$, and the double-\textbf{Q
}loop spin-vortex crystal phase. Finally, the ground states associated
with $\boldsymbol{\eta}_{3}$ are the single-\textbf{Q} stripe magnetic
phase with moments pointing out-of-plane, and the double-\textbf{Q
}charge-spin density-wave~\cite{allred16}. Note that these six states
are a subset of the states obtained from the minimization of Eq. (\ref{F_M}).
The restrictions arise from the symmetry properties that the spin
components must obey due to the finite spin-orbit coupling.

\subsection{Vestigial orders \label{subsec_vestigial}}

The vestigial orders in terms of the $\mathbf{M}_{a}$ order parameters
were previously discussed elsewhere by one of us~\cite{fernandes16}.
Here, we use group-theory to systematically derive all vestigial orders
in terms of the $\boldsymbol{\eta}_{\alpha}$ order parameters. This
corresponds to much more than a simple change of basis, as only the
$\boldsymbol{\eta}_{\alpha}$ vectors transform according to the actual
crystallographic space group. This will give rise to important qualitative
differences between the two cases. For instance, while certain vestigial
phases described by the free energy in Eq.~\eqref{F_M} break translational
symmetry and are associated with continuous order parameters, all
vestigial phases described by Eq.~\eqref{F_eta} preserve translational
symmetry and are associated with discrete order parameters.

A vestigial order parameter is a composite order parameter that can
order even in the absence of the primary order, and that breaks a
subset of the symmetries broken by the primary order parameter (for
a review, see Ref.~\onlinecite{fernandes18}). Vestigial order can
only appear if the primary order parameter transforms as a multi-dimensional
irrep, i.e. if the group is non-Abelian. In our case, we thus search
for bilinear combinations $\varphi_{\alpha}^{j}$ of the primary order
parameters $\boldsymbol{\eta}_{\alpha}$ that do not transform trivially
under the operations of the space group $P4/nmm$: 
\begin{equation}
\varphi_{\alpha}^{j}=\sum_{\mu,\nu}\eta_{\alpha}^{\mu}\Lambda_{\mu\nu}^{j}\eta_{\alpha}^{\nu}\,.\label{phi_j_general}
\end{equation}
Here, $\mu,\nu=1,\,2$ denote the two components of the vector $\boldsymbol{\eta}_{\alpha}$
and $\Lambda_{\mu\nu}^{j}$ is a $2\times2$ matrix. Since the three
$\boldsymbol{\eta}_{\alpha}$ belong to different irreps, and long-range
magnetic order develops only within a single irrep channel, we consider
only the composite order parameters associated with each irrep separately.
This is in agreement with previous RG calculations of Eq. (\ref{F_M})
in the presence of spin-orbit coupling~\cite{christensen18a,christensen18b}.

The $2\times2$ matrices $\Lambda_{\mu\nu}^{j}$ can be generally
expressed in terms of Pauli matrices $\boldsymbol{\tau}$. Their explicit
expressions are obtained by the decomposition of the products of irreps
$E_{Mi}$~\cite{cvetkovic13}: 
\begin{eqnarray}
E_{M1\left(M2\right)}\otimes E_{M1\left(M2\right)} & = & A_{1g}\oplus B_{2g}\oplus A_{2u}\oplus B_{1u}\\
E_{M3}\otimes E_{M3} & = & A_{1g}\oplus B_{2g}\oplus A_{1u}\oplus B_{2u}\,.\label{eq:irrep_product_m3}
\end{eqnarray}
The fact that both inversion-even ($g$) and inversion-odd ($u$)
one-dimensional irreps appear is because the two components of the
two-dimensional irreps transform oppositely under the glide-plane
symmetry. Note that, because $2\mathbf{Q}_{M}=(2\pi,\,2\pi)$, the
irreps on the right-hand side are those of the $P4/nmm$ space group
at the $\Gamma\equiv\left(0,0\right)$ point of the crystallographic
Brillouin zone, which is equivalent to the familiar point group $D_{4h}$.
Among those, it is not possible to form non-zero bilinears involving
$\boldsymbol{\eta}_{\alpha}$ that transform as $B_{1u}$ and $A_{1u}$.
Consequently, for each $\alpha$, there are only two possible composite
order parameters that transform non-trivially under $D_{4h}$ (i.e.
that do not transform as $A_{1g}$), namely: 
\begin{align}
\varphi_{\alpha}^{z} & \equiv\boldsymbol{\eta}_{\alpha}\tau^{z}\boldsymbol{\eta}_{\alpha}\,,\nonumber \\
\varphi_{\alpha}^{x} & \equiv\boldsymbol{\eta}_{\alpha}\tau^{x}\boldsymbol{\eta}_{\alpha}\,.\label{phi_j}
\end{align}
In terms of the original order parameters $\mathbf{M}_{a}$, we have:
\begin{align}
\varphi_{1}^{z}=M_{1,x}^{2}-M_{2,y}^{2}\,,\quad\varphi_{1}^{x}=2M_{1,x}M_{2,y}\,,\nonumber \\
\varphi_{2}^{z}=M_{1,y}^{2}-M_{2,x}^{2}\,,\quad\varphi_{2}^{x}=2M_{1,y}M_{2,x}\,,\nonumber \\
\varphi_{3}^{z}=M_{1,z}^{2}-M_{2,z}^{2}\,,\quad\varphi_{3}^{x}=2M_{1,z}M_{2,z}\,.\label{phi_j_aux}
\end{align}
All these composite order parameters have zero wave-vector, and thus
correspond to intra-unit cell order. In particular, $\varphi_{1}^{z}$,
$\varphi_{2}^{z}$, and $\varphi_{3}^{z}$ transform as the $B_{2g}$
irrep of $D_{4h}$; $\varphi_{1}^{x}$ and $\varphi_{2}^{x}$ transform
as $A_{2u}$; and $\varphi_{3}^{x}$ transform as $B_{2u}$. Because
order parameters that break the same symmetry must necessarily couple
bilinearly in the free energy expansion, they must either be all zero
or all non-zero at the same time. Therefore, we identify three distinct
vestigial phases, represented by the order parameters: $\Phi_{B_{2g}}\propto\varphi_{1}^{z}\propto\varphi_{3}^{z}\propto\varphi_{2}^{z}$;
$\Phi_{A_{2u}}\propto\varphi_{1}^{x}\propto\varphi_{2}^{x}$; $\Phi_{B_{2u}}\propto\varphi_{3}^{x}$.
All of them are Ising-like, zero wave-vector order parameters; the
subscript just indicates how they transform under the $D_{4h}$ group.
In the next section, we will see that they correspond to nematic order,
spin-current order, and checkerboard charge order, respectively. Interestingly,
the existence of three inter-related spin-driven Ising-nematic order
parameters, $\varphi_{1}^{z}$, $\varphi_{2}^{z}$, and $\varphi_{2}^{z}$
have been recently reported in an NMR study of a detwinned 122 iron
pnictide~\cite{curro18}.

Having established the vestigial order parameters, we now discuss
the actual vestigial phases. Formally, a vestigial phase is that in
which the symmetry-breaking composite order parameter is non-zero,
but the primary order parameter vanishes: 
\begin{equation}
\left\langle \boldsymbol{\eta}_{\alpha}\tau^{(x,z)}\boldsymbol{\eta}_{\alpha}\right\rangle \neq0\,,\quad\left\langle \boldsymbol{\eta}_{\alpha}\right\rangle =0\,.\label{eq:vestigial_general}
\end{equation}
Clearly, if $\left\langle \boldsymbol{\eta}_{\alpha}\right\rangle \neq0$,
it follows necessarily that at least one of the composite orders is
non-zero, $\left\langle \boldsymbol{\eta}_{\alpha}\tau^{(x,z)}\boldsymbol{\eta}_{\alpha}\right\rangle \neq0$.
The non-trivial aspect of the condition above arises from the fact
that the composite order can be finite even if the primary order is
not. Note also that $\left\langle \boldsymbol{\eta}_{\alpha}\boldsymbol{\tau}^{y}\boldsymbol{\eta}_{\alpha}\right\rangle $
is identically zero, as $\tau^{y}$ is antisymmetric, and that the
non-symmetry breaking composite combination $\left\langle \boldsymbol{\eta}_{\alpha}\tau^{0}\boldsymbol{\eta}_{\alpha}\right\rangle $
is always non-zero, as it transforms as the trivial $A_{1g}$ irrep
of $D_{4h}$. Physically, the latter corresponds simply to magnetic
amplitude fluctuations. To determine whether and when condition (\ref{eq:vestigial_general})
is satisfied, phenomenological considerations are not enough, and
one must resort to microscopic calculations, which will depend on
the model. Importantly, in our case, both the vestigial and primary
magnetic order parameters are Ising-like. Generally, it is possible
that the two Ising transitions are split and second- or first-order
(in which case there is a regime of vestigial phase) or simultaneous
and first-order (in which case there is no vestigial phase). Both
regimes have been theoretically seen, depending on model parameters~\cite{batista11,fernandes12}.

The outcome of our analysis is shown in Fig. \ref{fig:intertwined_schematic}.
It is remarkable that all these different broken-symmetry states arise
from the same primary degrees of freedom, which ultimately go back
to the six-dimensional super-vector $\boldsymbol{\mathcal{M}}$ in
Eq. (\ref{supervector}). We emphasize that each of the three classes
of magnetic order, represented by the irrep $E_{Mi}$ and shown in
Fig. \ref{fig:mag_order}, support two possible types of vestigial
order parameters: $\boldsymbol{\eta}_{\alpha}\tau^{z}\boldsymbol{\eta}_{\alpha}$,
associated with the single-\textbf{Q }phase, and $\boldsymbol{\eta}_{\alpha}\tau^{x}\boldsymbol{\eta}_{\alpha}$,
associated with the double-\textbf{Q }phase. While all three classes
support a $B_{2g}$ nematic vestigial phase, only the $E_{M1,M2}$
classes support an $A_{2u}$ spin-current vestigial states, whereas
only $E_{M3}$ supports a $B_{2u}$ charge-order vestigial phase.

\section{Orbital-order patterns in the vestigial phases}

\label{sec:el_impact}

Our analysis so far has focused exclusively on the spin degrees of
freedom. However, orbital degrees of freedom are essential to describe
the properties of the vestigial phases, particularly since it is the
very existence of a spin-orbit coupling that changes the form of the
magnetic free-energy from Eq.~\eqref{F_M} to Eq.~\eqref{F_eta}.

To proceed, we must first define what we mean by orbital degrees of
freedom, since the five $3d$ Fe orbitals are generally hybridized
throughout the Brillouin zone, and are thus not generically good quantum
numbers. However, if we focus on the low-energy states near the Fermi
level, we can restrict our analysis to the orbitals that give the
dominant spectral weight, namely, the $d_{xz}$, $d_{yz}$, and $d_{xy}$
orbitals. Moreover, because the Fermi pockets are typically small,
we can focus on the states close to the high-symmetry points of the
crystallographic Brillouin zone, i.e. the $\Gamma=(0,0)$ and $M=(\pi,\pi)$
points. Fortunately, at these high-symmetry points, we can define
pure orbital states. Thus, we can label the Bloch states near the
Fermi level by their orbital content, following the procedure outlined
in Ref.~\onlinecite{cvetkovic13}. Near the $\Gamma$ point, the
Fermi surface consists of two small hole pockets, and the relevant
orbitals are the $d_{xz}$ and $d_{yz}$, which transform as the $E_{g}$
representation of the $D_{4h}$ group. Thus, we define the electronic
doublet: 
\begin{eqnarray}
\Psi_{\Gamma,s}(\mbf{K})=\begin{pmatrix}d_{yz,s}(\mbf{k})\\
-d_{xz,s}(\mbf{k})
\end{pmatrix}\,,\label{eq:Psi_gamma}
\end{eqnarray}
in terms of the $xz$ and $yz$ orbital operators. Here, $s$ denotes
the spin quantum number. We use $\mbf{K}/\mathbf{X}$ to denote a
momentum in the crystallographic Brillouin zone/unit cell, while $\mbf{k}/\mathbf{x}$
corresponds to the 1-Fe Brillouin zone/unit cell (see Fig.~\ref{fig:unit_cell}).
Near the $M$ point, there are two small electron pockets whose orbital
contents are $d_{xz}/d_{yz}$ and $d_{xy}$. Because all irreps of
$P4/nmm$ at $M$ are doubly degenerate, we define two electronic
doublets: 
\begin{align}
\Psi_{M_{+},s}(\mbf{K}+\mbf{Q}_{M}) & =\begin{pmatrix}d_{xz,s}(\mbf{k}+\mbf{Q}_{2})\\
d_{xy,s}(\mbf{k}+\mbf{Q}_{1})
\end{pmatrix}\,,\nonumber \\
\Psi_{M_{-},s}(\mbf{K}+\mbf{Q}_{M}) & =\begin{pmatrix}d_{yz,s}(\mbf{k}+\mbf{Q}_{1})\\
d_{xy,s}(\mbf{k}+\mbf{Q}_{2})
\end{pmatrix}\,.\label{eq:Psi_M}
\end{align}
While the two elements of $\Psi_{M_{+},\sigma}$ transform as the
``upper'' elements of the $E_{M1}$ and $E_{M3}$ irreducible representations,
the two elements of $\Psi_{M_{-},\sigma}$ transform as the ``lower''
elements of $E_{M1}$ and $E_{M3}$~\cite{cvetkovic13}. The generic
non-interacting Hamiltonian in terms of these three doublet operators
was discussed in Ref.~\onlinecite{cvetkovic13}, and is shown again
in Appendix~\ref{app:kp_model}. Because its form is enforced only
by the symmetries of the $P4/nmm$ group, it can be used to fit any
experimentally determined band dispersion. Note that while the Bloch
states $\Psi_{\Gamma}$, $\Psi_{M_{\pm}}$ are defined in the crystallographic
Brillouin zone, the orbital operators $d_{xz}$, $d_{yz}$, $d_{xy}$
are defined in the 1-Fe Brillouin zone.

To obtain the orbital order patterns of each of the vestigial phases,
all we need now is to find the fermionic bilinears $\left\langle \Psi_{a}^{\dagger}\hat{\Lambda}\,\Psi_{b}\right\rangle $,
where $\hat{\Lambda}$ is a vertex in the space of electronic operators,
that transform according to the $B_{2g}$, $A_{2u}$, and $B_{2u}$
irreps of the $D_{4h}$ group. Since the vestigial orders have zero
wave-vector, we look for combinations $\left\langle \Psi_{\Gamma}^{\dagger}\hat{\Lambda}\,\Psi_{\Gamma}\right\rangle $
and $\left\langle \Psi_{M_{\pm}}^{\dagger}\hat{\Lambda}\,\Psi_{M_{\pm}}\right\rangle $.
Using the product tables for the $P4/nmm$ group and including the
role of the spin operators, we obtain the orbital order parameters
detailed below. For the full derivation, see Appendix \ref{app:fermionic_terms}.

\subsection{$B_{2g}$ nematic order}

We start with the case of the $\Phi_{B_{2g}}$ vestigial order, which
is the vestigial phase associated with the $E_{M_{1}}$, $E_{M_{2}},$
and $E_{M_{3}}$ single-\textbf{Q} stripe magnetic phase, as discussed
in Sec. \ref{sec:vestigial_phases}. There are three different types
of orbital order parameters that appear concomitantly with the condensation
of $\Phi_{B_{2g}}$ (see Ref.~\onlinecite{fernandes14b}): 
\begin{align}
\Delta_{B_{2g}}^{(1)} & =\left\langle d_{yz,s}^{\dagger}(\mbf{k})d_{yz,s}(\mbf{k})-d_{xz,s}^{\dagger}(\mbf{k})d_{xz,s}(\mbf{k})\right\rangle \label{eq:Delta_1_B2g}\\
\Delta_{B_{2g}}^{(2)} & =\left\langle d_{yz,s}^{\dagger}(\mbf{k}+\mbf{Q}_{1})d_{yz,s}(\mbf{k}+\mbf{Q}_{1})\right.\nonumber \\
 & \left.-d_{xz,s}^{\dagger}(\mbf{k}+\mbf{Q}_{2})d_{xz,s}(\mbf{k}+\mbf{Q}_{2})\right\rangle \label{eq:Delta_2_B2g}\\
\Delta_{B_{2g}}^{(3)} & =\left\langle d_{xy,s}^{\dagger}(\mbf{k}+\mbf{Q}_{1})d_{xy,s}(\mbf{k}+\mbf{Q}_{1})\right.\nonumber \\
 & \left.-d_{xy,s}^{\dagger}(\mbf{k}+\mbf{Q}_{2})d_{xy,s}(\mbf{k}+\mbf{Q}_{2})\right\rangle \,.\label{eq:Delta_3_B2g}
\end{align}
Note that we have omitted the momentum summations for convenience.
The first two terms can be expressed as combinations of onsite ferro-orbital
order and bond-orbital order involving the $d_{xz}$ and $d_{yz}$
orbitals, whereas the third term corresponds to bond orbital order
involving the $d_{xy}$ orbital. The breaking of tetragonal symmetry
warrants identifying this vestigial phase as a nematic phase. We emphasize
that this orbital pattern is rather different than a uniform occupation
of the $d_{xz}$ and $d_{yz}$ orbitals, as it has previously been
pointed out theoretically in \cite{yamakawa16,chubukov16,Benfatto16}
and observed experimentally in Refs.~\onlinecite{Susuki15,Ding15}.

\begin{figure}[!h]
\centering
\includegraphics[width=0.96\columnwidth]{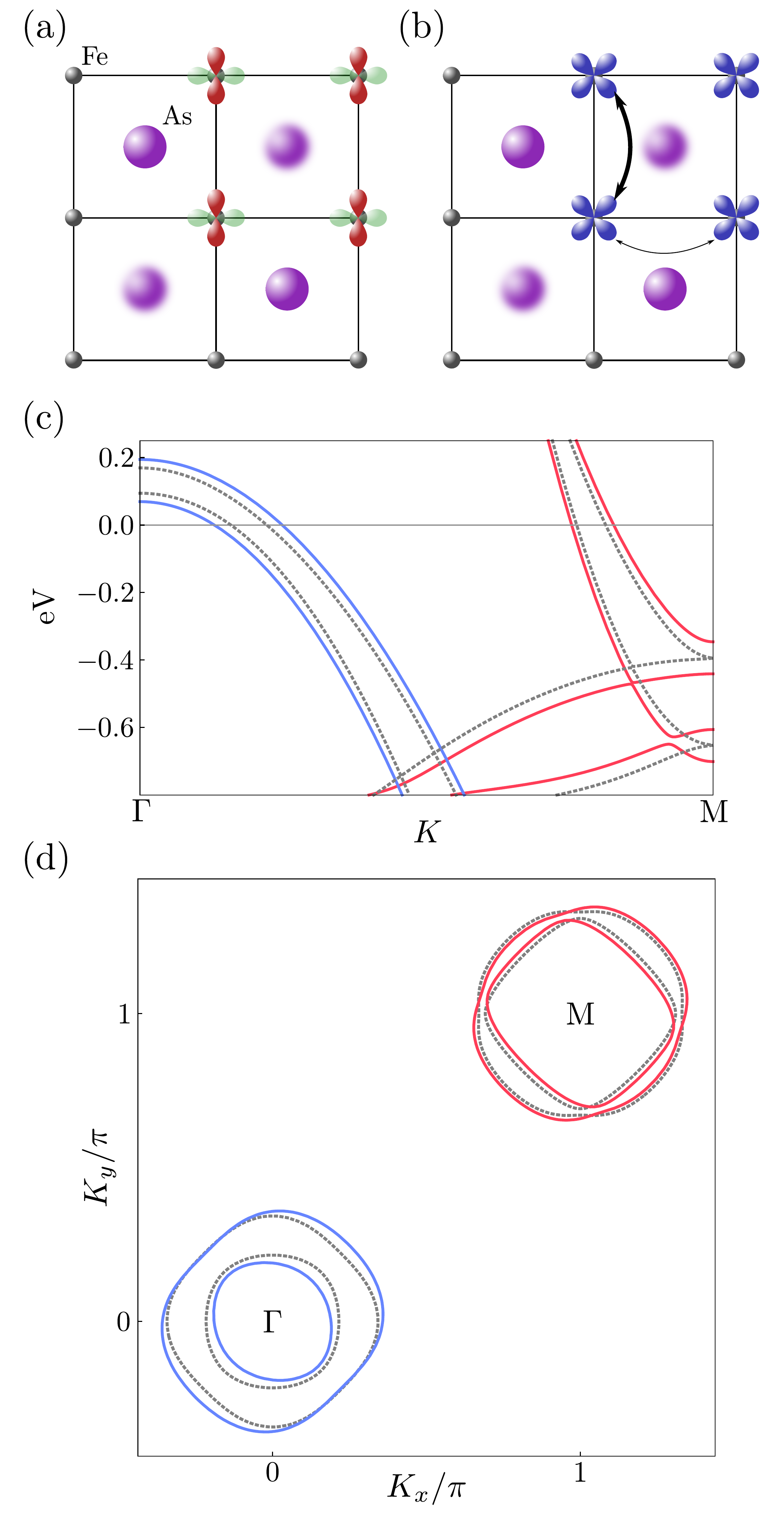}
\caption{\label{fig:b2g} Effects of $B_{2g}$ vestigial nematic order. In
(a) we illustrate the ferro-orbital order involving the $d_{xz}$
and $d_{yz}$ Fe $3d$-orbitals ($\Delta_{B_{2g}}^{(1)}\protect\neq0$)
while in (b), the bond-nematic order (hopping anisotropy) involving
the $d_{xy}$ orbitals ($\Delta_{B_{2g}}^{(3)}\protect\neq0$). The band structure and Fermi surface in the vestigial phase are shown in (c) and (d), respectively. The dotted lines show the band structure in the non-ordered phase. Note the Pomeranchuk-type distortion of the Fermi surface. The parameters used here are given in Appendix \ref{app:kp_model}.}
\end{figure}

The real-space orbital order patterns are illustrated in Figs. \ref{fig:b2g}(a)
and (b). While the former represents an onsite $d_{xz}/d_{yz}$ ferro-orbital
order, the latter represents a hopping anisotropy between nearest-neighbors
$d_{xy}$ orbitals (i.e. bond order). In Figs.~\ref{fig:b2g}(c)
and (d), we show the impact of these orbital order parameters on the
band dispersions and on the Fermi surface. The main effects are the
well-known splitting of the energy doublets at the $M$ point and
the Pomeranchuk-like distortion of the Fermi pockets, which are no
longer $C_{4}$ symmetric. In these figures, we also included the
finite atomic spin-orbit coupling, which lifts the degeneracy of the
energy doublet at $\Gamma$ and hybridizes the two electron pockets
already in the non-nematic phase~\cite{borisenko16,fernandes14b}.

\begin{figure}
\centering \includegraphics[width=0.96\columnwidth]{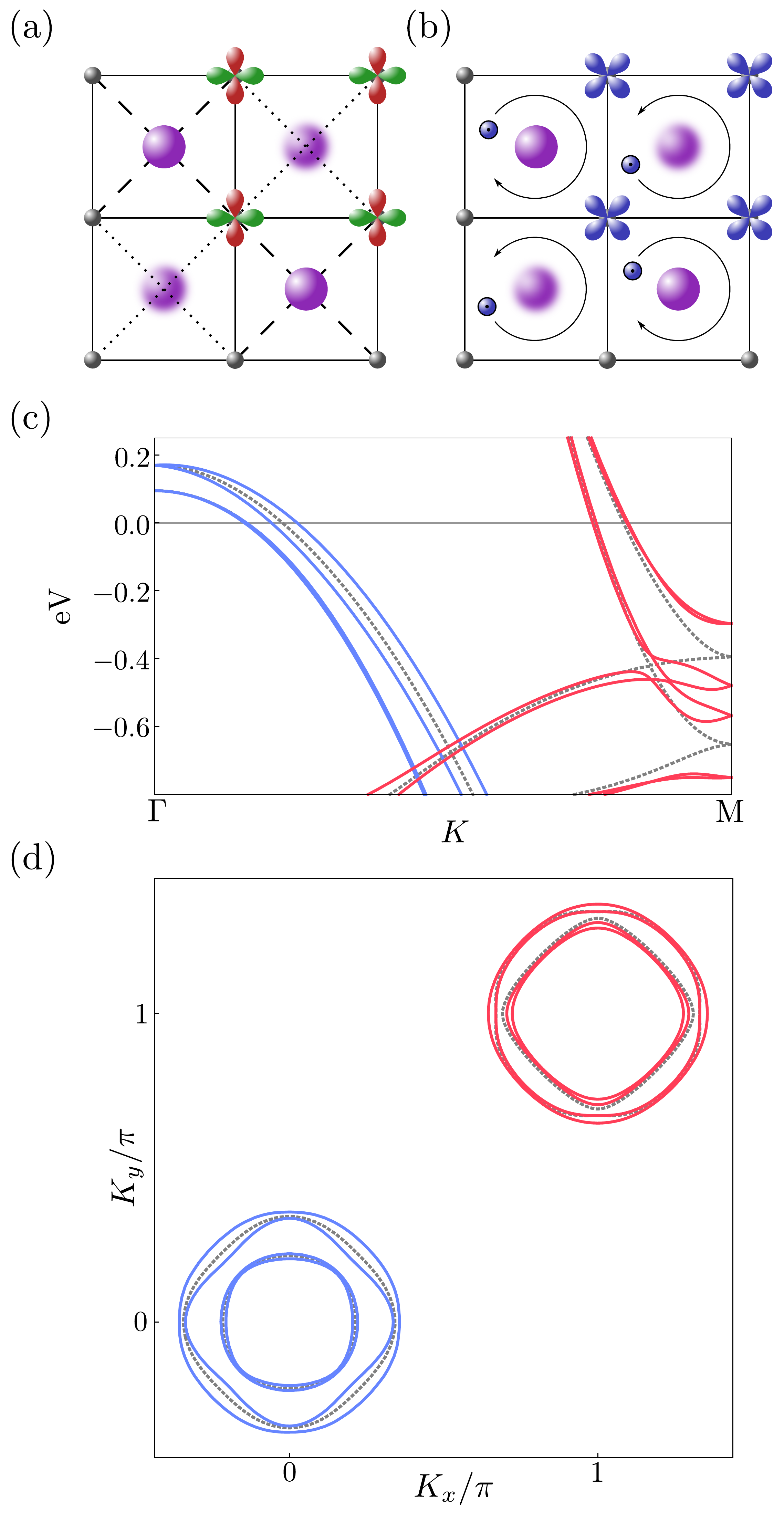}
\caption{\label{fig:a2u} Effects of $A_{2u}$ vestigial spin-current order. In (a), the bonds between next-nearest-neighbor Fe-atoms become inequivalent in a staggered pattern as a result of $\Delta_{A_{2u}}^{(1)}\protect\neq0$. This renders the As/Se atoms above and below the Fe-plane inequivalent as well. In (b), we illustrate the formation of staggered spin-currents involving the $d_{xy}$ orbitals ($\Delta_{A_{2u}}^{(2)}\protect\neq0$). The band structure and Fermi surface in the vestigial phase are shown
in (c) and (d), respectively. Note the doubling of the number of bands
due to the lifting of spin degeneracy. The parameters used here are given in Appendix \ref{app:kp_model}.}
\end{figure}

\subsection{$A_{2u}$ spin-current order}

Let us now consider the vestigial phase with $\Phi_{A_{2u}}$ order
parameter, which is associated with the $E_{M_{1}}$, $E_{M_{2}}$
double-\textbf{Q} spin-vortex crystal phase discussed in Sec. \ref{sec:vestigial_phases}.
There are two momentum-independent fermionic bilinears that transform
as $A_{2u}$; both involve only states at the $M$ point: 
\begin{align}
\Delta_{A_{2u}}^{(1)} & =\left\langle d_{xz,s}^{\dagger}(\mbf{k}+\mbf{Q}_{2})d_{yz,s}(\mbf{k}+\mbf{Q}_{1})+\mathrm{h.c.}\right\rangle \label{eq:Delta_1_A2u}\\
\Delta_{A_{2u}}^{(2)} & =i\left\langle d_{xy,s}^{\dagger}(\mbf{k}+\mbf{Q}_{2})\sigma_{ss'}^{z}d_{xy,s'}(\mbf{k}+\mbf{Q}_{1})-\mathrm{h.c.}\right\rangle \,.\label{eq:Delta_2_A2u}
\end{align}
The first term corresponds to an intra-unit cell staggered deformation
of the bonds connecting next-nearest-neighbor Fe atoms, and as such
renders the positions of the As/Se atoms above and below the Fe plane
inequivalent \cite{Agterberg17}. The second term corresponds to an
intra-unit cell staggered spin-current, polarized along the $z$ axis,
involving only the $d_{xy}$ orbitals \textendash{} hence the identification
of this phase as spin-current order. Both types of order are illustrated,
in real space, in Figs.~\ref{fig:a2u}(a) and (b), respectively.
Additional spin-dependent terms are discussed in Appendix \ref{app:fermionic_terms}.

Although, at first sight, it may seem that the states near $\Gamma$
are not affected by $\Phi_{A_{2u}}$, this only happens if we restrict
the analysis to momentum-independent fermionic bilinears. Allowing
for terms that are linear in momentum, we find several orbital-order
combinations that are reminiscent of the Rashba- and Dresselhaus-type
terms typically found in systems that break inversion symmetry. For
instance, two of these terms are: 
\begin{align}
\Delta_{A_{2u}}^{(3)} & =\left\langle \left(k_{x}\sigma_{ss'}^{y}-k_{y}\sigma_{ss'}^{x}\right)\right.\nonumber \\
 & \left.\left(d_{xz,s}^{\dagger}(\mbf{k})d_{xz,s'}(\mbf{k})+d_{yz,s}^{\dagger}(\mbf{k})d_{yz,s'}(\mbf{k})\right)\right\rangle \\
\Delta_{A_{2u}}^{(5)} & =\left\langle \left(k_{x}\sigma_{ss'}^{x}-k_{y}\sigma_{ss'}^{y}\right)\right.\nonumber \\
 & \left.\left(d_{xz,s}^{\dagger}(\mbf{k})d_{yz,s'}(\mbf{k})+d_{yz,s}^{\dagger}(\mbf{k})d_{xz,s'}(\mbf{k})\right)\right\rangle \,.\label{eq:Delta_3_A2u}
\end{align}
Recall that $k_{x}$ and $k_{y}$ refer to momenta in the 1-Fe Brillouin
zone and $\sigma^{x}$ and $\sigma^{y}$ refer to spins along the
Fe-Fe bonds. All four allowed linear-in-momentum terms are shown explicitly
in Appendix~\ref{app:fermionic_terms}.

In Figs.~\ref{fig:a2u}(c) and (d), we show the electronic band dispersion
reconstructed by all these orbital-order patterns. Besides the splitting
of the energy doublets at $M$, which is similar to the case of the
nematic vestigial order, the most salient feature is the doubling
of the number of bands, despite the absence of translational or rotational
symmetry breaking. This doubling is a consequence of the fact that
the bands are no longer spin-degenerate due to the combination of
a finite spin-orbit coupling and an order parameter that breaks inversion
symmetry (i.e. it transforms as $A_{2u}$). Indeed, such a band splitting
is commonly found in systems where the explicit breaking of inversion
symmetry (e.g. by an applied electric field) induces a Rashba coupling.
What is unique to the $A_{2u}$ vestigial phase is that the Rashba-type
coupling is the result of an spontaneous symmetry-breaking. Being
a composite spin order parameter, its magnitude is set by the strength
of magnetic fluctuations, which could in principle be sizable near
a magnetic critical point. It is important to emphasize that the Fermi
surface remains symmetric under a $90^{\circ}$ rotation and under
an in-plane inversion. This is a consequence of the fact that the
vestigial order parameter transforms as $A_{2u}$.

\begin{figure}
\centering \includegraphics[width=0.96\columnwidth]{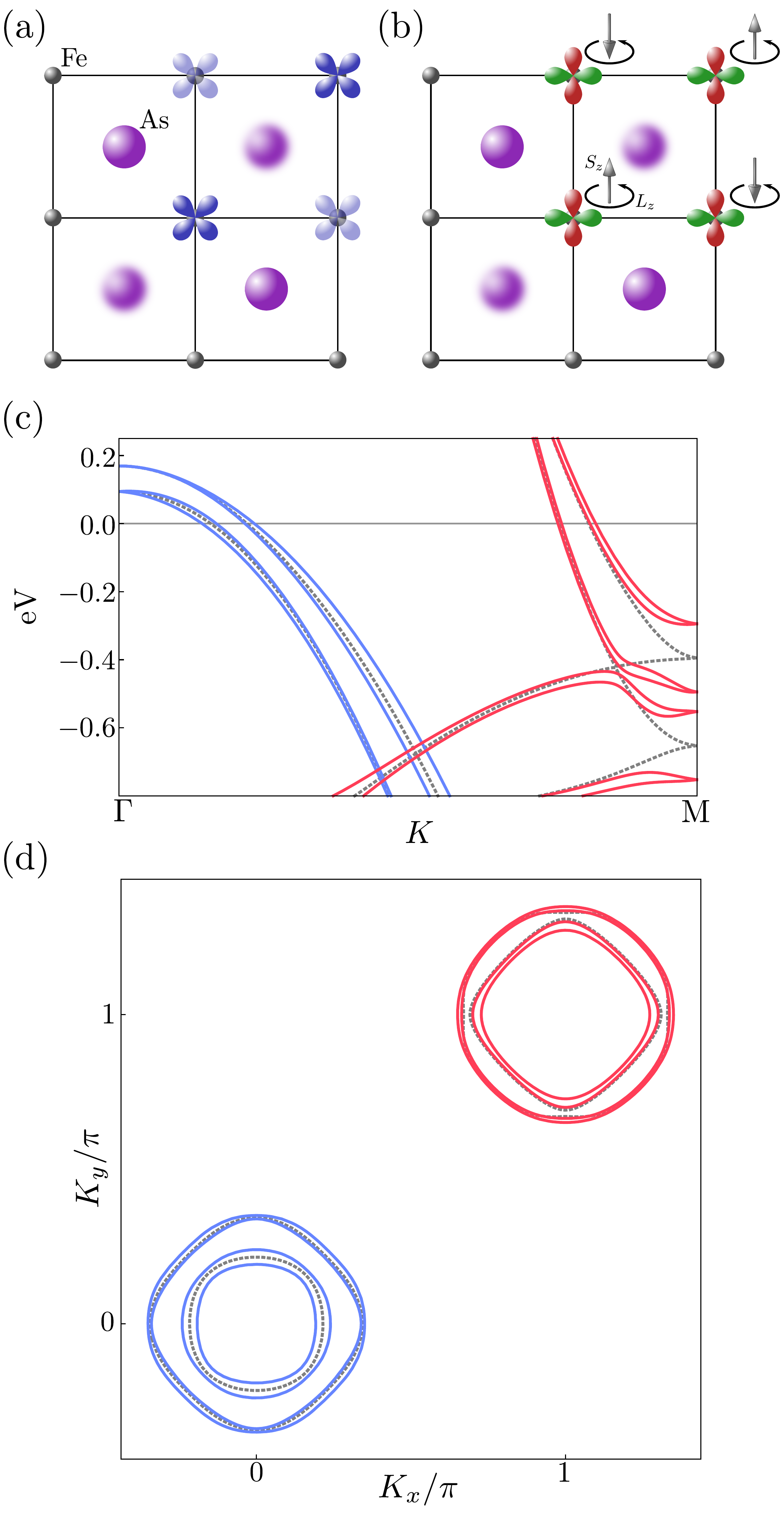}
\caption{\label{fig:b2u} Effects of the $B_{2u}$ vestigial charge order.
In (a) we show the induced charge order in the $d_{xy}$ orbital ($\Delta_{B_{2u}}^{(1)}\protect\neq0$)
and in (b) the emergence of a staggered $L_{z}S_{z}$ spin-orbital
coupling between the $d_{xz}$ and $d_{yz}$ orbitals ($\Delta_{B_{2u}}^{(2)}\protect\neq0$).
The band structure and Fermi surfaces in the vestigial phase are shown
in (c) and (d), respectively. Note the doubling of the number of bands
due to the lifting of spin degeneracy. The parameters used here are
given in Appendix \ref{app:kp_model}.}
\end{figure}

\subsection{$B_{2u}$ charge order}

Finally, we analyze the orbital order patterns associated with the
vestigial order with $\Phi_{B_{2u}}$ order parameter, which is related
to the $E_{M_{3}}$ double-\textbf{Q} charge-spin density-wave phase.
There are two momentum-independent fermionic bilinears transforming
as $B_{2u}$: 
\begin{align}
\Delta_{B_{2u}}^{(1)} & =\left\langle d_{xy,s}^{\dagger}(\mbf{k}+\mbf{Q}_{2})d_{xy,s}(\mbf{k}+\mbf{Q}_{1})+\text{h.c.}\right\rangle \\
\Delta_{B_{2u}}^{(2)} & =i\left\langle d_{xz,s}^{\dagger}(\mbf{k}+\mbf{Q}_{2})\sigma_{ss'}^{z}d_{yz,s'}(\mbf{k}+\mbf{Q}_{1})-\text{h.c.}\right\rangle \,.\label{eq:Delta_2_B2u}
\end{align}
The first term corresponds to an occupation imbalance of the $d_{xy}$
orbital on neighboring sites, as illustrated in Fig.~\ref{fig:b2u}(a).
Note that this is an intra-unit cell order. The second term corresponds
to an intra-unit cell staggered $L_{z}S_{z}$-type spin-orbit coupling,
illustrated in Fig.~\ref{fig:b2u}(b). Appendix \ref{app:fermionic_terms}
shows additional spin-dependent terms are discussed.

As with the case of $A_{2u}$ vestigial order, electronic states near
the $\Gamma$ point are affected by terms linear in momentum. As explained
above, these terms resemble the Rashba- and Dresselhaus-type spin-orbit
couplings commonly found in systems that break inversion symmetry.
In our case however, the multi-orbital structure allows for multiple
inequivalent terms: 
\begin{align}
\Delta_{B_{2u}}^{(3)} & =\left\langle \left(k_{x}\sigma_{ss'}^{x}-k_{y}\sigma_{ss'}^{y}\right)\right.\nonumber \\
 & \left.\left(d_{xz,s}^{\dagger}(\mbf{k})d_{xz,s'}(\mbf{k})+d_{yz,s}^{\dagger}(\mbf{k})d_{yz,s'}(\mbf{k})\right)\right\rangle \\
\Delta_{B_{2u}}^{(5)} & =\left\langle \left(k_{x}\sigma_{ss'}^{y}-k_{y}\sigma_{ss'}^{x}\right)\right.\nonumber \\
 & \left.\left(d_{xz,s}^{\dagger}(\mbf{k})d_{yz,s'}(\mbf{k})+d_{yz,s}^{\dagger}(\mbf{k})d_{xz,s'}(\mbf{k})\right)\right\rangle \,.\label{eq:Delta_3_B2u}
\end{align}
All the induced spin-orbit terms are listed in Appendix~\ref{app:fermionic_terms}.
The reconstruction of the electronic dispersion in the vestigial phase
that displays $B_{2u}$ order is shown in Figs.~\ref{fig:b2u}(c)
and (d). As with the $\Phi_{A_{2u}}$ case, the spin-degeneracy of
the bands is lifted and translational symmetry remains unbroken. The
doublets at $M$ are split in a manner reminiscent of the $B_{2g}$
nematic case, although the Fermi surface remains symmetric under a
$90^{\circ}$ rotation since the order parameter transforms as $B_{2u}$.

\section{Experimental Consequences}

\label{sec:ind_orders}

The symmetry classification above reveals the existence of three different
types of vestigial phases associated with the magnetic ground states
of the iron pnictides, which trigger unique orbital order patterns.
All these vestigial order parameters are Ising-like (i.e. scalar)
that have zero in-plane wave-vector. As such, they transform as different
irreducible representations of the $P4/nmm$ group at the $\Gamma$
point, which coincide with the irreps of the $D_{4h}$ space group.
In particular, we found a vestigial nematic $B_{2g}$ phase, a vestigial
spin-current $A_{2u}$ phase, and a vestigial charge-ordered $B_{2u}$
phase. In this section, we discuss different experimental manifestations
of these vestigial orders based on their coupling to non-electronic
degrees of freedom, such as the lattice, as well as to external electromagnetic
fields.

\subsection{Coupling to the lattice}\label{sec:latt_coupling}

In the long wavelength limit, the elastic degrees of freedom are described
in terms of derivatives of the displacement vector $\mathbf{U}$.
For a tetragonal lattice, there are six independent strain modes.
Two of them transform as $A_{1g}$ and do not change the symmetry
of the lattice (e.g. volume collapse); two of them involve out-of-plane
shear distortions and transform as $E_{g}$; and another two involve
only in-plane lattice distortions and transform as $B_{1g}$ and $B_{2g}$.
These last two correspond to the strain fields $\varepsilon_{B_{1g}}=\partial_{X}U_{X}-\partial_{Y}U_{Y}$
and $\varepsilon_{B_{2g}}=\partial_{X}U_{Y}+\partial_{Y}U_{X}$, respectively.
Note that $\mathbf{U}(\mathbf{X})$ is defined in the crystallographic
unit cell.

A simple symmetry analysis reveals that the only vestigial order parameter
that couples linearly to a symmetry-breaking strain mode is the nematic
one, $\Phi_{B_{2g}}$, which couples to the shear mode $\varepsilon_{B_{2g}}$.
Such a linear coupling drives a softening of the sound velocity of
the acoustic phonon mode that propagates along the $[100]$ direction.
This coupling has been widely discussed in the literature and explored
to experimentally probe both long-range nematic order and nematic
fluctuations~\cite{fernandes10,yoshisawa12,chu12,kuo16,kretzschmar16,bohmer16,kaneko17,weber18}.
Because $\Phi_{B_{2u}}$ and $\Phi_{A_{2u}}$ do not couple linearly
to any of the strain modes, the lattice remains tetragonal in the
spin-current and charge-ordered vestigial phases.

Besides the elastic lattice modes, associated with acoustic phonons,
the system also has optical phonon modes. Among those, there is an
$A_{2u}$ zone-center optical phonon, with an energy of about $\Omega_{A_{2u}}\approx40$
meV, corresponding to an in-phase displacement $Z$ of the Fe and
As/Se atoms along the out-of-plane axis \cite{Tacon17}. Symmetry
dictates that there must be a linear coupling between this $Z$ displacement
and the vestigial order parameter $\Phi_{A_{2u}}$. Therefore, the
condensation of $\Phi_{A_{2u}}$ leads to a softening of the $A_{2u}$
optical phonon. Although the crystal structure remains tetragonal
after $\Omega_{A_{2u}}\rightarrow0$, this linear coupling opens the
interesting possibility of probing fluctuations associated with the
spin-current vestigial phase by inelastic neutron scattering measurements
of the phonon spectrum. Furthermore, since this phonon mode is infrared
active, it is in principle possible to manipulate the spin-current
order parameter via pump-probe spectroscopy.

We finish this section by noting that the vestigial order parameter
$\Phi_{B_{2u}}$ does not couple linearly to either acoustic or optical
phonon modes. This implies that the onset of charge-ordered vestigial
phase is not accompanied by any changes in the atomic positions of
the crystal structure.

\subsection{Conjugate fields and transverse fields}

The analysis above reveals that the shear strain $\varepsilon_{B_{2g}}$
acts as a conjugate field to the vestigial nematic order parameter
$\Phi_{B_{2g}}$. Thus, controlled application of uniaxial strain,
for instance via piezoelectric devices, offers a unique route to probe
nematicity. This has been routinely employed to study nematicity in
the iron pnictides and other systems that display nematic order.

An interesting question is whether there are experimentally accessible
conjugate fields to the other vestigial order parameters obtained
here, $\Phi_{A_{2u}}$ and $\Phi_{B_{2u}}$. Interestingly, an electric
field applied along the $z$ axis, $E_{z}$, also transforms as $A_{2u}$,
and thus behaves as a conjugate field to the spin-current order parameter
$\Phi_{A_{2u}}$. Such a field can be experimentally applied in a
controlled way via electrostatic gating of thin films. It can also
appear in a less controlled way due to the existence of a substrate,
as is the case in FeSe grown on SrTiO$_{3}$~\cite{yan12,he13,tan13,shen14,ge15,huang17},
or due to a particular feature of the crystalline structure, as is
the case of the 1144 pnictides~\cite{iyo16,meier18}. The linear
coupling between $E_{z}$ and $\Phi_{A_{2u}}$ opens up the unexplored
possibility of experimentally probing spin-current order and spin-current
fluctuations in a systematic way.

It is also interesting to note that the order parameters $\Phi_{A_{2u}}$
and $\Phi_{B_{2g}}$ are not completely unrelated, as they live in
the same two-dimensional subspace spanned by the two-component magnetic
order parameter $\boldsymbol{\eta}_{1}$ (or $\boldsymbol{\eta}_{2}$).
This is a direct consequence of the common vestigial nature of both
orders, and manifested clearly in Eq. (\ref{phi_j}): while $\Phi_{B_{2g}}$
is associated with the $\tau^{z}$ Pauli matrix living in the ``Bloch
sphere'' set by the the two components of $\boldsymbol{\eta}_{1}$,
$\Phi_{A_{2u}}$ is described in terms of the $\tau^{x}$ Pauli matrix.
In view of the standard commutation relations between Pauli matrices,
it is expected that one of the order parameters acts effectively as
a ``transverse field'' to the other one. Of course, this by itself
does not imply that non-trivial Berry phases appear in the free energy
(for an analysis related to the simpler case of orbital nematicity,
see Ref.~\onlinecite{maharaj17}). Yet, it does suggest that the
conjugate field to $\Phi_{A_{2u}}$ will have an effect on $\Phi_{B_{2g}}$
(and vice-versa), as they live in the same subspace. Thus, an electric
field $E_{z}$ may be used to tune the nematic transition, instead
of shear strain. The advantage is that the former does not explicitly
break the symmetry that is spontaneously broken by $\Phi_{B_{2g}}$.
This effect, which we dub \emph{electro-nematic}, is unique to the
vestigial nature of the nematic order, and would not be present if
nematicity was simply a manifestation of ferro-orbital order, see
Ref.~\onlinecite{maharaj17}.

As for the vestigial order parameter $\Phi_{B_{2u}}$, we did not
identify a simple experimentally accessible quantity that could act
as its conjugate field. It is possible, however, to combine two different
quantities whose product transforms as $B_{2u}$. This is achieved
for instance by combining a biaxial strain $\varepsilon_{B_{1g}}$
and a perpendicular electric field $E_{z}$. Since $B_{1g}\otimes A_{2u}=B_{2u}$,
there is a trilinear coupling between $\Phi_{B_{2u}}$, $\varepsilon_{B_{1g}}$,
and $E_{z}$. Of course, the disadvantage of this trilinear coupling
is that $E_{z}$ also induces $\Phi_{A_{2u}}$.

\subsection{Intertwinment with Néel antiferromagnetic order \label{subsec_Neel}}

The magnetically ordered states discussed in our work, described in
terms of the order parameters $\boldsymbol{\eta}_{\alpha}$, are associated
with the ordering vectors $\left(\pi,0\right)$ and $\left(0,\pi\right)$
of the Fe-only Brillouin zone. Another type of magnetic order that
has been observed in certain hole-doped iron pnictides is the well-known
Néel order, described by an order parameter $\mathbf{N}$, and associated
with the ordering vector $\left(\pi,\pi\right)$ of the Fe-only Brillouin
zone \cite{Tucker13}. Of course, in the crystallographic Brillouin
zone, $\left(\pi,\pi\right)$ is folded onto $(0,0)$. But the key
point is that, because $\left(\pi,\pi\right)=\left(\pi,0\right)+(0,\pi)$,
one expects that $\mathbf{N}$ will couple to a bilinear combination
of $\eta_{\alpha}^{\mu}\eta_{\alpha}^{\nu}$, with $\mu\neq\nu$.
Those are precisely the vestigial order parameters defined in Eq.
(\ref{phi_j}).

This simple observation motivates us to investigate the coupling between
the Néel order parameter $\mathbf{N}$ and the vestigial order parameters
$\Phi$. To ensure time-reversal invariance, since $\mathbf{N}$ is
odd under time-reversal whereas $\Phi$ is even, this coupling can
only take place in the presence of an external magnetic field $\mathbf{H}$.
Note that $\mathbf{H}$ is a pseudo-vector whose out-of-plane component
transforms as $A_{2g}$ and whose in-plane components transform as
$E_{g}$ \cite{cvetkovic13}. In contrast, for the Néel order on the
Fe-sites, $\mbf{N}$, $(N_{x},-N_{y})^{T}$ forms an $E_{u}$ doublet
while $N_{z}$ transforms as $B_{1u}$. A symmetry analysis reveals
that there is no trilinear coupling involving $\mathbf{H}$, $\mathbf{N}$,
and the vestigial nematic order parameter $\Phi_{B_{2g}}$.

\begin{figure}
\centering \includegraphics[width=0.95\columnwidth]{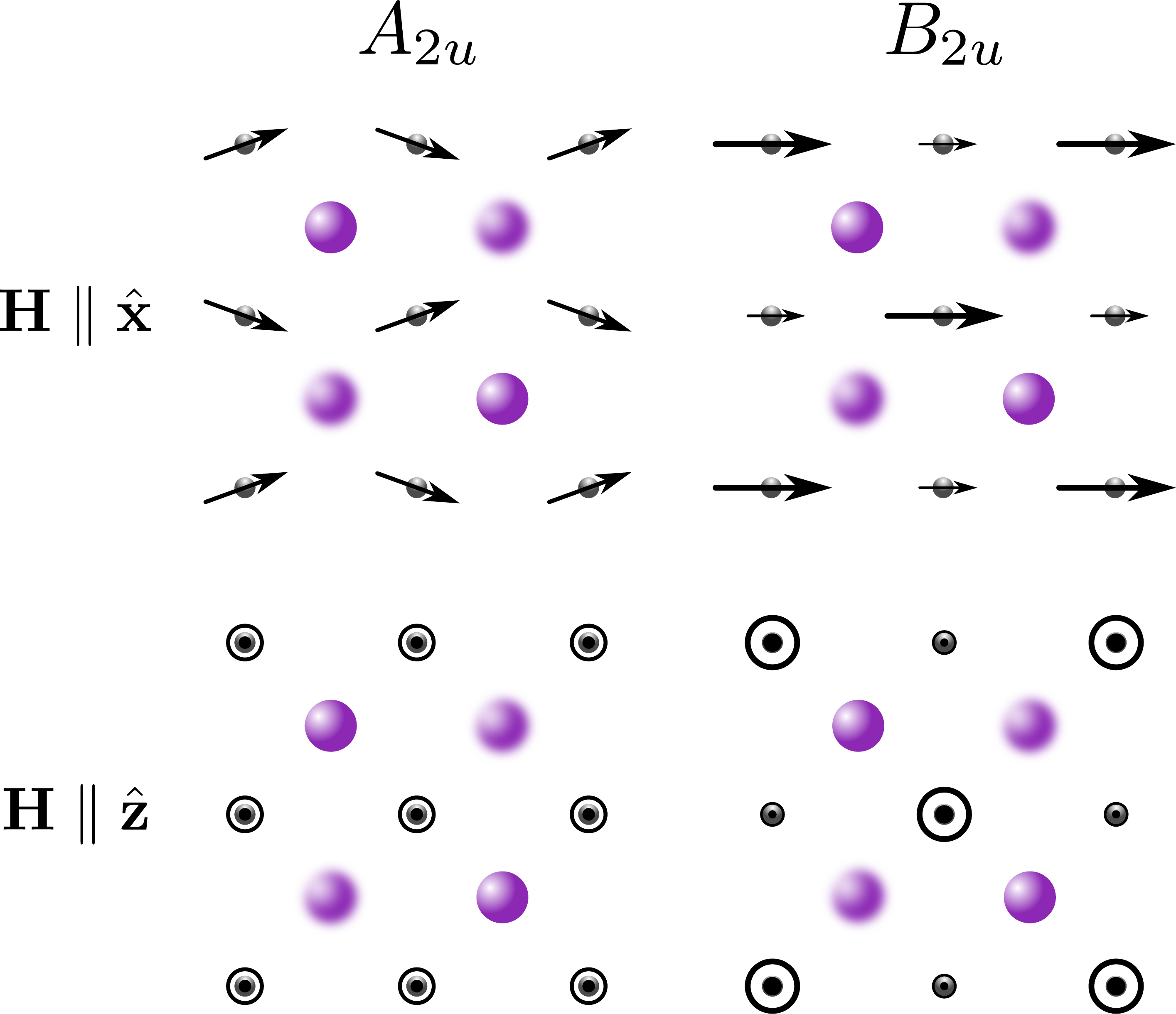}
\caption{\label{fig:induced_order} Illustration of the induced magnetic configurations
in the $A_{2u}$ and $B_{2u}$ paramagnetic vestigial phases induced
by the presence of an external magnetic field $\mbf{H}$. Besides
the usually induced ferromagnetic order, Néel order is also induced
in this case, resulting in ferrimagnetic or canted antiferromagnetic
configurations. For out-of-plane fields, currents are induced around
the As/Se atoms. These are omitted in the figure.}
\end{figure}

The situation is different for the vestigial order parameter $\Phi_{B_{2u}}$.
The free energy in such a case allows for a trilinear coupling of
the form:

\begin{eqnarray}
\mathcal{F}_{B_{2u}} & = & \lambda_{B_{2u}}\Phi_{B_{2u}}\mbf{H}\cdot\mbf{N}\,,\label{eq:induced_order_B2u}
\end{eqnarray}
where $\lambda_{B_{2u}}$ is a coupling constant. The physical implication
of this term is clear: in the vestigial charge-orderd phase, where
$\Phi_{B_{2u}}\neq0$, application of a uniform magnetic field induces
Néel order polarized parallel to the field direction. Because $\mathbf{H}$
inevitably induces ferromagnetic order, the resulting magnetic configuration
is \emph{ferrimagnetic}, as illustrated in Fig.~\ref{fig:induced_order}.

A trilinear coupling is also allowed in the case of the vestigial
order parameter $\Phi_{A_{2u}}$, although its form is different than
that of Eq. (\ref{eq:induced_order_B2u}):

\begin{eqnarray}
\mathcal{F}_{A_{2u}} & = & \lambda_{A_{2u}}\Phi_{A_{2u}}\left(H_{x}N_{y}+H_{y}N_{x}\right)\,,\label{eq:induced_order_a2u}
\end{eqnarray}
where $\lambda_{A_{2u}}$ denotes a coupling constant. Here, the subscripts
$x,\,y$ refer to the 1-Fe unit cell, i.e. they are parallel to the
Fe-Fe nearest-neighbor directions. In the spin-current phase, where
$\Phi_{A_{2u}}\neq0$, the polarization of the Néel order induced
by the external field depends crucially on its direction. An out-of-plane
field, for instance, does not induce any Néel order, as illustrated
in Fig.~\ref{fig:induced_order}. On the other hand, for $\mathbf{H}\parallel\hat{\mathbf{x}}$
or $\mathbf{H}\parallel\hat{\mathbf{y}}$, the Néel order parameter
is polarized along the perpendicular in-plane direction, $\mathbf{N}\parallel\hat{\mathbf{y}}$
or $\mathbf{N}\parallel\hat{\mathbf{x}}$, respectively. Due to the
additional parallel ferromagnetic component induced by $\mathbf{H}$,
the resulting magnetic configuration is that of a \emph{canted antiferromagnet
}(see Fig.~\ref{fig:induced_order}). The relative angle $\Delta\theta$
between the ferromagnetic and Néel components that are induced by
an in-plane magnetic field applied along a direction that makes an
angle $\theta_{\mathrm{H}}$ with the $x$ axis is given by:

\[
\Delta\theta=2\theta_{\mathrm{H}}\pm\frac{\pi}{2}
\]
where the $\pm$ sign depends on the sign of the coupling constant
$\lambda_{A_{2u}}$. Clearly, for $\theta_{\mathrm{H}}=\pm\frac{\pi}{4}$,
corresponding to a field applied along the (anti)diagonal, there is
no canting and the induced order is ferrimagnetic, similarly to the
case with $\Phi_{B_{2u}}$ order.

We dub the rather unusual effect described by Eqs. (\ref{eq:induced_order_B2u})-(\ref{eq:induced_order_a2u})
of Néel order being induced by a uniform magnetic field the \emph{ferro-Néel
effect. }Its existence highlights the non-trivial paramagnetic character
of the vestigial phases. It provides yet another possible route to
experimentally probe the vestigial spin-current and charge-ordered
phases. The key manifestation, which can be probed by neutron scattering
experiments, would be the emergence of a magnetic Bragg peak at $\left(\pi,\pi\right)$
(of the 1-Fe Brillouin zone) upon application of a magnetic field.

\section{Impact of three-dimensionality: contrasting 122 and 1111, 111, 11
compounds}

\label{sec:diff_11_122}

The results derived in this work refer to the space group $P4/nmm$,
for which the crystallographic unit cell only contains a single FeAs/Se
layer. Because the materials are layered compounds, the stacking pattern
of the layers is crucial to determine the applicability of our results
to the different families of iron-based superconductors. In the case
where the layers are simply stacked on top of each other, as is the
case in the 1111, 111, and 11 compounds (such as LaFeAsO, NaFeAs,
and FeSe, respectively), the space group is indeed $P4/nmm$. Consequently,
all results derived here apply directly to these systems. Specifically,
the vestigial ordered states have ordering vectors $\mathbf{Q}=\left(0,0,0\right)$
in the crystallographic Brillouin zone. This also implies that the
Néel order parameter of Sec. \ref{subsec_Neel} has an ordering vector
$\left(\pi,\pi,0\right)$ in the 1-Fe Brillouin zone.

The situation changes qualitatively in the case of the 122 compounds
(e.g. BaFe$_{2}$As$_{2}$), where the layers are stacked in a staggered
pattern. In this case, the appropriate space group is not the non-symmorphic
$P4/nmm$ group but instead the symmorphic $I4/mmm$. While a full
group-theoretical analysis is beyond the scope of this work, we can
still determine the main qualitative changes in our results in this
case. This can be accomplished by starting from the artificial 1-Fe
Brillouin zone and noting that the main difference between the compounds
with $P4/nmm$ and $I4/mmm$ space groups is on the $z$-component
of the momentum vector $\mathbf{P}_{\mathrm{fold}}$ responsible for
folding the larger 1-Fe Brillouin zone onto the crystallographic 2-Fe
Brillouin zone~\cite{ku11,valenti14,fernandes16b}. In particular,
for the compounds with $P4/nmm$ space group, the folding vector is
$\mathbf{P}_{\mathrm{fold}}=\left(\pi,\pi,0\right)$ whereas for the
compounds with $I4/nmm$ space group, it is $\mathbf{P}_{\mathrm{fold}}=\left(\pi,\pi,\pi\right)$.
Using these folding vectors, the wave-vector in the crystallographic
unit cell can be written as $\mathbf{Q}=\hat{\mbf{R}}_{\pi/4}\cdot(\mathbf{q}+\mathbf{P}_{\mathrm{fold}})$,
where $\mathbf{q}$ is the wave-vector in the 1-Fe Brillouin zone
and $\hat{\mbf{R}}_{\pi/4}$ realizes a $45^{\circ}$ rotation around
the $\hat{\mbf{z}}$-axis (see Fig.~\ref{fig:unit_cell}).

Because the vestigial nematic order parameter $\Phi_{B_{2g}}$ has
ordering vector $\mathbf{q}=\left(0,0,0\right)$ in the 1-Fe Brillouin
zone, it remains a zero wave-vector order in the 2-Fe Brillouin zone
for both types of systems. Thus, in regards to the nematic phase,
the results are the same for both $P4/nmm$ and $I4/mmm$ compounds.
On the other hand, the 1-Fe Brillouin zone ordering vector of the
two other vestigial order parameters $\Phi_{A_{2u}}$ and $\Phi_{B_{2u}}$
is $\mathbf{q}=\left(\pi,\pi,0\right)$. This means that, in the 2-Fe
(crystallographic) Brillouin zone, the ordering vector is $\mathbf{Q}=\left(0,0,0\right)$
for the $P4/nmm$ compounds, but $\mathbf{Q}=\left(0,0,\pi\right)$
for the $I4/mmm$ compounds. Consequently, these vestigial orders
break translational symmetry by doubling the periodicity along the
$z$ direction. While this property provides another experimentally
accessible way of detecting these vestigial orders (e.g. via X-ray
scattering), it also implies that $E_{z}$ is not a conjugate field
for $\Phi_{A_{2u}}$ and that $A_{2u}$ spin-current order is not
accompanied by a soft zone-center phonon. The results for the coupling
to the Néel order parameter of Sec.~\ref{subsec_Neel} remain the
same, since the ordering vector of latter in the 1-Fe Brillouin zone
is $\left(\pi,\pi,0\right)$.

We finish this section by emphasizing that the premise of our framework
is that the dominant magnetic fluctuations are peaked at $\left(\pi,0,\pi\right)$
and $\left(0,\pi,\pi\right)$ in the 1-Fe Brillouin zone, i.e. they
are stripe-like magnetic fluctuations. Although this is the situation
in the majority of iron-based materials, this is not the case for
all of them. The prime example is the 11 compound FeTe, which displays
double-stripe magnetic order, with ordering vector $\left(\frac{\pi}{2},\frac{\pi}{2},\pi\right)$
in the 1-Fe Brillouin zone~\cite{li09,bao09,Dagotto17}. While our
results are not relevant for that compound, this type of magnetic
order does support its own set of vestigial orders, as discussed in Ref.
\onlinecite{Zhang17}. These vestigial orders have been invoked to explain
the phase diagram of a related Ti-based oxypnictide~\cite{Zhang17}.
We also emphasize that, in FeSe and related intercalated systems,
even though the dominant magnetic fluctuations are stripe-type~\cite{Boothroyd15,Zhao16},
it remains widely debated whether the nematic phase is a vestigial
order or the consequence of a separate instability~\cite{baek15,glasbrenner15,DHLee15,chubukov16,Benfatto16,yamakawa16}. The main evidence against a simple vestigial scenario is that the low-energy magnetic fluctuations, as measured by NMR, are rather weak above the onset of nematic order [cite Bohmer]. Typically, vestigial order is triggered by the correlation length of the primary order reaching an often large threshold value~\cite{fernandes18}. Moreover, the fact that upon application of pressure the nematic transition temperature is suppressed whereas the magnetic transition temperature is enhanced~\cite{bohmer18} suggests a scenario of competing phases rather than intertwined vestigial phases. However, as shown by the first-principle calculations of Ref.~\onlinecite{glasbrenner15}, the magnetic ground state of FeSe seems to be much more degenerate than $(\pi,0)$ and $(0,\pi)$ (in the 1Fe BZ). Instead, there is a family of nearly degenerate magnetic states with different ordering vectors. What they have in common is that they all support a vestigial nematic phase, despite directly competing for different magnetic configurations. This suggests an interesting case of vestigial order being stabilized over a wider temperature range by multiple competing magnetic instabilities; as such, this scenario certainly deserves further investigation.

\section{Discussion and conclusions}

\label{sec:conclusions}

In this paper we presented a simple, yet powerful framework that provides
a unifying description of the complexity of the underdoped phase diagram
of the iron pnictides. Such a framework consists both of primary phases
that display long-range magnetic order and of their vestigial phases
that intertwine spin and orbital degrees of freedom. A plethora of
electronic orders that break different symmetries of the FeAs/Se layer
while displaying comparable transition temperatures emerge within
this framework, from stripe magnetic order and nematicity to $C_{4}$
magnetism, charge order, and loop spin-current order. Formally, all
these states arise from the condensation of the symmetry-allowed simple
and composite order parameters formed by the six components of the
super-vector $\boldsymbol{\mathcal{M}}$ in Eq. (\ref{supervector}),
as schematically shown in Fig. \ref{fig:intertwined_schematic}.

Whereas previous works focusing on the idealized Fe-only square lattice
have identified some of these phases, all electronic orders discussed
here respect the symmetries of the crystallographic non-symmorphic
space group that describes the FeAs/Se layer. These symmetries qualitatively
change the nature of the vestigial phases, and the orbital-order patterns
that accompany the composite spin orders. Besides the widely investigated
Ising-nematic order parameter $\Phi_{B_{2g}}$, which transforms as
the $B_{2g}$ irrep of the tetragonal group, two additional Ising-like,
zero wave-vector composite order parameters are found, transforming
as the $A_{2u}$ and $B_{2u}$ irreps (denoted by $\Phi_{A_{2u}}$
and $\Phi_{B_{2u}}$).

While $\Phi_{B_{2g}}$ is associated with a combination of onsite-
and bond-orbital orders involving the $d_{xz}$, $d_{yz}$, $d_{xy}$
orbitals, $\Phi_{A_{2u}}$ is accompanied by spin-current order involving
the $d_{xy}$ orbitals and staggered $d_{xz}/d_{yz}$ orbital hybridization.
Similarly, $\Phi_{B_{2u}}$ is accompanied by checkerboard charge-order
related to the $d_{xy}$ orbitals and staggered $d_{xz}/d_{yz}$ spin-orbit
coupling. Moreover, the onset of $\Phi_{A_{2u}}$ and $\Phi_{B_{2u}}$
also triggers different types of Rashba-like and Dresselhaus-like
spin-orbit couplings involving the $d_{xz}/d_{yz}$ orbitals. Importantly,
these spin-orbit coupling terms are a consequence of a spontaneous
symmetry breaking driven by magnetic fluctuations, rather than an
explicit broken symmetry due to e.g. a substrate or an applied electric
field.

These rather unusual orbital patterns lead to a variety of interesting
effects unveiled in our work: the lifting of the spin-degeneracy of
the electronic band structure in the vestigial $A_{2u}$ and $B_{2u}$
phases; the electro-nematic effect, by which an electric field applied
perpendicular to the FeAs/Se layer acts effectively as a transverse
field to the nematic order parameter; and the ferro-Néel effect, describing
the fact that Néel order is induced by a uniform magnetic field in
the vestigial $A_{2u}$ and $B_{2u}$ phases, resulting in ferrimagnetic
and canted antiferromagnetic spin configurations.

The most promising iron-based compounds where these unusual $A_{2u}$
and $B_{2u}$ vestigial phases may be realized are those for which
the primary $C_{4}$ magnetic phases have been observed. Interestingly,
not only doping, but also pressure, have been shown to be capable
of tuning the magnetic ground state from stripe to $C_{4}$~\cite{hassinger16,bohmer18}.
Recently, indirect evidence for charge-order, presumably due to a
non-zero $\Phi_{B_{2u}}$, has been reported in ARPES experiments
in hole-doped compounds~\cite{yi18}. Although the experiment was
performed inside the charge-spin density-wave phase, it would be interesting
to investigate whether the effect persists into the paramagnetic phase.
In the 1144 compound, it has been argued that a non-zero $\Phi_{A_{2u}}$
exists due to a small symmetry-breaking field associated with the
crystal structure of this material, combined with proximity to a spin-vortex
magnetic phase~\cite{meier18}.

More broadly, the formalism discussed here could also be relevant
to other correlated systems that have the same space group as the
iron pnictides and display magnetic or nematic orders. One interesting
case is that of URu$_{2}$Si$_{2}$, which has the same $I4/mmm$
space group as the 122 pnictides. Nematic order has been suggested
below the hidden-order temperature by certain experiments, which may
be an instability on its own right or a $B_{2g}$ vestigial phase
\cite{Okazaki11,Riggs15}. Other experiments indicate non-magnetic
Fermi surface folding with ordering vector $\left(0,0,\pi\right)$
in the hidden-order phase, which one may speculate to be related to
a $A_{2u}$ or $B_{2u}$ vestigial order~\cite{hassinger10,meng13}.
Another potentially relevant $5f$-electron system is the compound
CeAuSb$_{2}$, whose space group $P4/nmm$ is the same as the 11,
111, and 1111 pnictides. Recent neutron scattering experiments revealed
a change in the magnetic ground state from single-\textbf{Q }stripe
to double-\textbf{Q }magnetic order as function of an applied magnetic
field~\cite{marcus18}. Whether vestigial phases also emerge in this
phase diagram remains to be established.
\begin{acknowledgments}
The authors are grateful to W. R. Meier for inspiring discussions
on the group-theoretical description of iron pnictides. The authors
further acknowledge fruitful discussions with C. Batista, E. Berg,
P. Canfield, A. Chubukov, S. Kivelson, A. Kreyssig, P. Orth, J. Schmalian,
and R. Valent{í}. This work was supported by the U.S. Department
of Energy, Office of Science, Basic Energy Sciences, under Award number
DE-SC0012336. R. M. F. also acknowledges support from the Research Corporation for Science Advancement via the Cottrell Scholar Award.
\end{acknowledgments}

\appendix

\section{Low-energy electronic model}

\label{app:kp_model}

For completeness, we here provide the details of the non-interacting
Hamiltonian used to obtain the band structures shown in Figs.~\ref{fig:b2g}\textendash \ref{fig:b2u}.
They follow the parametrization first proposed in Ref. \onlinecite{cvetkovic13}.
Note that the results presented in the main text are independent on
the specific form of the Hamiltonian, since all the arguments are
based on symmetry. The basis used here and throughout the paper is
\begin{eqnarray}
\Psi_{s}(\mbf{K})=\begin{pmatrix}\Psi_{M_{+},s}(\mbf{K}+\mbf{Q}_{M})\\
\Psi_{M_{-},s}(\mbf{K}+\mbf{Q}_{M})\\
\Psi_{\Gamma,s}(\mbf{K})
\end{pmatrix}\,,\label{eq:psi}
\end{eqnarray}
where $\Psi_{\Gamma,s}$ and $\Psi_{M,s}$ are given in Eqs.~\eqref{eq:Psi_gamma}
and \eqref{eq:Psi_M}. The Hamiltonian is given by: 
\begin{eqnarray}
\mathcal{H} & = & \begin{pmatrix}h_{+}(\mbf{K}) & h_{+-}^{{\rm SOC}} & 0\\
(h_{+-}^{{\rm SOC}})^{\dagger} & h_{-}(\mbf{K}) & 0\\
0 & 0 & h_{\Gamma}(\mbf{K})+h_{\Gamma}^{{\rm SOC}}
\end{pmatrix}\,.\label{eq:ham}
\end{eqnarray}
Recall that $\mbf{K}$ labels momentum in the crystallographic 2-Fe
Brillouin zone and $\mbf{Q}_{M}=(\pi,\pi)$. $\mathcal{H}$ is a $12\times12$
matrix with components 
\begin{eqnarray}
 &  & h_{\pm}(\mbf{K})=\nonumber \\
 &  & \begin{pmatrix}\epsilon_{1}+\frac{\mbf{K}^{2}}{2m_{1}}\pm a_{1}K_{x}K_{y} & -iv_{\pm}(\mbf{K})\\
iv_{\pm}(\mbf{K}) & \epsilon_{3}+\frac{\mbf{K}^{2}}{2m_{3}}\pm a_{3}K_{x}K_{y}
\end{pmatrix}\otimes\sigma^{0}\,,\nonumber \\
 &  & h_{\Gamma}(\mbf{K})=\nonumber \\
 &  & \begin{pmatrix}\epsilon_{\Gamma}+\frac{\mbf{K}^{2}}{2m_{\Gamma}}+bK_{x}K_{y} & c(K_{x}^{2}-K_{y}^{2})\\
c(K_{x}^{2}-K_{y}^{2}) & \epsilon_{\Gamma}+\frac{\mbf{K}^{2}}{2m_{\Gamma}}-bK_{x}K_{y}
\end{pmatrix}\otimes\sigma^{0}\,,
\end{eqnarray}
where 
\begin{eqnarray}
v_{\pm}(\mbf{K}) & = & v(\pm K_{x}+K_{y})+p_{1}(\pm K_{X}^{3}+K_{y}^{3})\nonumber \\
 & + & p_{2}K_{x}K_{y}(K_{x}\pm K_{y})\,,
\end{eqnarray}
and the SOC components are 
\begin{eqnarray}
h_{\Gamma}^{{\rm SOC}} & = & \frac{1}{2}\lambda_{\Gamma}\begin{pmatrix}0 & -i\\
i & 0
\end{pmatrix}\otimes\sigma^{3}\,,\nonumber \\
h_{+-}^{{\rm SOC}} & = & \frac{i}{2}\lambda_{M}\left[\begin{pmatrix}0 & 1\\
0 & 0
\end{pmatrix}\otimes\sigma^{1}+\begin{pmatrix}0 & 0\\
1 & 0
\end{pmatrix}\otimes\sigma^{2}\right]\,.
\end{eqnarray}
The parameters can be determined on a case-by-case basis by fitting
to tight-binding results. For concreteness, we adopted the ones provided
in Ref.~\onlinecite{cvetkovic13}, based on the band structure of
Ref.~\onlinecite{cvetkovic09}. For the plots in Figs. \ref{fig:b2g},
\ref{fig:a2u}, and \ref{fig:b2u}, we set $\lambda_{\Gamma}=\lambda_{M}=75{\rm {meV}}$
to make the SOC effects more pronounced. For the momentum independent
fermionic bilinears we set $\Delta_{B_{2g}}^{(1,2)}=\Delta_{A_{2u}}^{(1,2)}=\Delta_{B_{2u}}^{(1,2)}=100{\rm meV}$
to ensure a more pronounced effect. For the Rashba- and Dresselhaus-like
SOC terms, we set $\Delta_{A_{2u}}^{(3,4,5,6)}=\Delta_{B_{2u}}^{(3,4,5,6)}=10{\rm meV}$.

\section{Fermionic bilinears transforming as $A_{2u}$ and $B_{2u}$}

\label{app:fermionic_terms}

Here we present the fermionic bilinears constructed from the electronic
operator (\ref{eq:psi}) that transform as the $A_{2u}$ and $B_{2u}$
irreps, and are thus induced by the onset of $\Phi_{A_{2u}}$ and
$\Phi_{B_{2u}}$. We use Eqs. (\ref{eq:Psi_gamma}) and (\ref{eq:Psi_M})
to express the bilinears in terms of the orbital operators with momentum
defined in the 1-Fe Brillouin zone. Note that the bilinears that transform
as $B_{2g}$ have been previously discussed in Ref.~\onlinecite{fernandes14b}.

\subsection{Electron pockets}

We first focus on bilinears of the form $\left\langle \Psi_{M_{\pm}}^{\dagger}\hat{\Lambda}\,\Psi_{M_{\pm}}\right\rangle $,
which involve states at the electron pockets. The vertex $\hat{\Lambda}$
has spin, orbital and momentum parts, $\hat{\Lambda}=\hat{\Lambda}^{(s)}\otimes\hat{\Lambda}^{(o)}\otimes\hat{\Lambda}^{(k)}$.
In the case of spin- and momentum-independent bilinears, $\hat{\Lambda}^{(s)}$
and $\hat{\Lambda}^{(k)}$ are identity matrices and the orbital vertex
$\hat{\Lambda}^{(o)}$ must transform according to the $A_{2u}$ and
$B_{2u}$ irreps. Using the vertices tabulated in Ref.~\onlinecite{cvetkovic13},
we find: 
\begin{align}
\Delta_{A_{2u}}^{(1)} & =\left\langle d_{xz,s}^{\dagger}(\mbf{k}+\mbf{Q}_{2})d_{yz,s}(\mbf{k}+\mbf{Q}_{1})\right\rangle \,.\nonumber \\
\Delta_{B_{2u}}^{(1)} & =\left\langle d_{xz,s}^{\dagger}(\mbf{k}+\mbf{Q}_{2})d_{yz,s}(\mbf{k}+\mbf{Q}_{1})\right\rangle \,.
\end{align}
In this expression and hereafter, sums over momentum and spin indices
are left implicit. In addition to these terms, that are also bilinears
that are not spin diagonal. Consider first the case $\hat{\Lambda}^{(s)}=\sigma^{z}$,
discussed in the main text. Since $\sigma^{z}$ transforms as $A_{2g}$,
all we need is to find the orbital vertices $\hat{\Lambda}^{(o)}$
that transform as $A_{1u}$ and $B_{1u}$, since $A_{2g}\otimes A_{1u}=A_{2u}$
and $A_{2g}\otimes B_{1u}=B_{2u}$. We obtain: 
\begin{align}
\Delta_{A_{2u}}^{(2)} & =i\left\langle d_{xy,s}^{\dagger}(\mbf{k}+\mbf{Q}_{2})\sigma_{ss'}^{z}d_{xy,s'}(\mbf{k}+\mbf{Q}_{1})-\mathrm{h.c.}\right\rangle \,,\nonumber \\
\Delta_{B_{2u}}^{(2)} & =i\left\langle d_{xz,s}^{\dagger}(\mbf{k}+\mbf{Q}_{2})\sigma_{ss'}^{z}d_{yz,s'}(\mbf{k}+\mbf{Q}_{1})-\text{h.c.}\right\rangle \,.
\end{align}
Note that the imaginary prefactor $i$ ensures that time-reversal
symmetry is preserved. We can also construct momentum-independent
bilinears with the spin-vertex $\hat{\Lambda}^{(s)}=(\sigma^{x},\sigma^{y})$,
which transforms as $E_{g}$, and orbital vertices $\hat{\Lambda}^{(o)}\sim E_u$. The result is
\begin{eqnarray}
	\Delta^{(7)}_{A_{2u}} &=& i \left\langle d^{\dagger}_{xy,s}(\mbf{k}+\mbf{Q}_2)\sigma^y_{ss'}d_{xz,s'}(\mbf{k}+\mbf{Q}_2) - \text{h.c.} \right\rangle \nonumber \\
	&-& i\left\langle d^{\dagger}_{xy,s}(\mbf{k}+\mbf{Q}_1)\sigma^x_{ss'}d_{yz,s'}(\mbf{k}+\mbf{Q}_1) - \text{h.c.} \right\rangle \nonumber \\
	\Delta^{(7)}_{B_{2u}} &=& i \left\langle d^{\dagger}_{xy,s}(\mbf{k}+\mbf{Q}_2)\sigma^x_{ss'}d_{xz,s'}(\mbf{k}+\mbf{Q}_2) - \text{h.c.} \right\rangle \nonumber \\
	&-& i\left\langle d^{\dagger}_{xy,s}(\mbf{k}+\mbf{Q}_1)\sigma^y_{ss'}d_{yz,s'}(\mbf{k}+\mbf{Q}_1) - \text{h.c.} \right\rangle\,. \nonumber \\
\end{eqnarray}
These terms are very similar to the regular SOC terms that are present even in the absence of vestigial order. However, the momentum dependence differs, which results in a staggered SOC, similar to $\Delta^{(2)}_{A_{2u}}$.

\subsection{Hole pockets}

We now turn to the bilinears involving states at the hole pockets,
which have the form $\left\langle \Psi_{\Gamma}^{\dagger}\hat{\Lambda}\,\Psi_{\Gamma}\right\rangle $.
As before, we write the vertex as $\hat{\Lambda}=\hat{\Lambda}^{(s)}\otimes\hat{\Lambda}^{(o)}\otimes\hat{\Lambda}^{(k)}$.
The doublet $\Psi_{\Gamma}$, corresponding to $d_{xz}$ and $d_{yz}$
orbitals, transforms as the $E_{g}$ irrep of the $P4/nmm$ group
at $\Gamma$. Thus, the momentum-independent bilinears constructed
from such fermions will transform as one of the \emph{gerade} irreps:
\begin{eqnarray}
E_{g}\otimes E_{g}=A_{1g}\otimes A_{2g}\otimes B_{1g}\otimes B_{2g}\,.\label{aux1}
\end{eqnarray}
To construct fermionic bilinears transforming as \emph{ungerade} irreps
it is necessary to consider vertices $\hat{\Lambda}^{(s)}$ and $\hat{\Lambda}^{(k)}$
that are odd under inversion. Since spin-vertices are even under inversion,
the vertices that are odd under inversion must be $\hat{\Lambda}^{(k)}$,
which must be odd in momentum. Hence, to preserve time-reversal symmetry,
the spin vertices $\hat{\Lambda}^{(s)}$ therefore cannot be the identity.
We thus construct the allowed vertices by combining $(k_{x},k_{y})$,
which transforms as $E_{u}$, and $(\sigma^{x},\sigma^{y})$, which
transforms as $E_{g}$: 
\begin{eqnarray}
E_{u}\otimes E_{g} & = & A_{1u}\otimes A_{2u}\otimes B_{1u}\otimes B_{2u}\,.\label{aux2}
\end{eqnarray}
Combining Eqs. (\ref{aux1}) and (\ref{aux2}) gives: 
\begin{eqnarray}
E_{g}\otimes E_{g}\otimes E_{u}\otimes E_{g}=4A_{1u}\otimes4A_{2u}\otimes4B_{1u}\otimes4B_{2u}\,,\nonumber \\
\end{eqnarray}
The numerical prefactors denote how many copies of the respective
irreps can be constructed. The $4$ different $A_{2u}$ terms result
from 
\begin{eqnarray}
A_{1g} & \otimes & A_{2u}\nonumber \\
A_{2g} & \otimes & A_{1u}\nonumber \\
B_{1g} & \otimes & B_{2u}\nonumber \\
B_{2g} & \otimes & B_{1u}\,,\label{eq:a1g_a2u}
\end{eqnarray}
whereas the $4$ different $B_{2u}$ terms are: 
\begin{eqnarray}
A_{1g} & \otimes & B_{2u}\nonumber \\
A_{2g} & \otimes & B_{1u}\nonumber \\
B_{1g} & \otimes & A_{2u}\nonumber \\
B_{2g} & \otimes & A_{1u}\,.\label{eq:b2g_a1u}
\end{eqnarray}
In the expressions above, the \emph{gerade} irreps originate from
the bilinear $\Psi_{\Gamma}^{\dagger}\Psi_{\Gamma}$, Eq.~(\ref{aux1}),
whereas the \emph{ungerade} irreps come $\hat{\Lambda}=\hat{\Lambda}^{(s)}\otimes\hat{\Lambda}^{(k)}$,
Eq.~(\ref{aux2}). Writing down these combinations explicitly, we
have 
\begin{eqnarray}
\left[A_{1g}\right] & \sim & d_{xz,s}^{\dagger}(\mbf{k})d_{xz,s'}(\mbf{k})+d_{yz,s}^{\dagger}(\mbf{k})d_{yz,s'}(\mbf{k})\nonumber \\
\left[A_{2g}\right] & \sim & d_{xz,s}^{\dagger}(\mbf{k})d_{yz,s'}(\mbf{k})-d_{yz,s}^{\dagger}(\mbf{k})d_{xz,s'}(\mbf{k})\nonumber \\
\left[B_{1g}\right] & \sim & d_{xz,s}^{\dagger}(\mbf{k})d_{yz,s'}(\mbf{k})+d_{yz,s}^{\dagger}(\mbf{k})d_{xz,s'}(\mbf{k})\nonumber \\
\left[B_{2g}\right] & \sim & d_{xz,s}^{\dagger}(\mbf{k})d_{xz,s'}(\mbf{k})-d_{yz,s}^{\dagger}(\mbf{k})d_{yz,s'}(\mbf{k})\,.
\end{eqnarray}
Moreover: 
\begin{eqnarray}
\left[A_{1u}\right] & \sim & k_{x}\sigma_{ss'}^{x}+k_{y}\sigma_{ss'}^{y}\nonumber \\
\left[A_{2u}\right] & \sim & k_{x}\sigma_{ss'}^{y}-k_{y}\sigma_{ss'}^{x}\nonumber \\
\left[B_{1u}\right] & \sim & k_{x}\sigma_{ss'}^{y}+k_{y}\sigma_{ss'}^{x}\nonumber \\
\left[B_{2u}\right] & \sim & k_{x}\sigma_{ss'}^{x}-k_{y}\sigma_{ss'}^{y}\,.
\end{eqnarray}
In the expressions above, the orbitals, spins, and momenta, are defined
with respect to the coordinate system of the single Fe atom square
lattice. Combining these expressions according to Eqs.~\eqref{eq:a1g_a2u}
and \eqref{eq:b2g_a1u} gives: 
\begin{eqnarray}
\Delta_{A_{2u}}^{(3)} & \sim & \big\langle\left(k_{x}\sigma_{ss'}^{y}-k_{y}\sigma_{ss'}^{x}\right)\nonumber \\
 &  & \qquad\left(d_{xz,s}^{\dagger}(\mbf{k})d_{xz,s'}(\mbf{k})+d_{yz,s}^{\dagger}(\mbf{k})d_{yz,s'}(\mbf{k})\right)\big\rangle\nonumber \\
\Delta_{A_{2u}}^{(4)} & \sim & \big\langle\left(k_{x}\sigma_{ss'}^{x}+k_{y}\sigma_{ss'}^{y}\right)\nonumber \\
 &  & \qquad\left(d_{xz,s}^{\dagger}(\mbf{k})d_{yz,s'}(\mbf{k})-d_{yz,s}^{\dagger}(\mbf{k})d_{xz,s'}(\mbf{k})\right)\big\rangle\nonumber \\
\Delta_{A_{2u}}^{(5)} & \sim & \big\langle\left(k_{x}\sigma_{ss'}^{x}-k_{y}\sigma_{ss'}^{y}\right)\nonumber \\
 &  & \qquad\left(d_{xz,s}^{\dagger}(\mbf{k})d_{yz,s'}(\mbf{k})+d_{yz,s}^{\dagger}(\mbf{k})d_{xz,s'}(\mbf{k})\right)\big\rangle\nonumber \\
\Delta_{A_{2u}}^{(6)} & \sim & \big\langle\left(k_{x}\sigma_{ss'}^{y}+k_{y}\sigma_{ss'}^{x}\right)\nonumber \\
 &  & \qquad\left(d_{xz,s}^{\dagger}(\mbf{k})d_{xz,s'}(\mbf{k})-d_{yz,s}^{\dagger}(\mbf{k})d_{yz,s'}(\mbf{k})\right)\big\rangle\nonumber \\
\label{eq:b2g_b1u_full}
\end{eqnarray}
for $A_{2u}$ and 
\begin{eqnarray}
\Delta_{B_{2u}}^{(3)} & \sim & \big\langle\left(k_{x}\sigma_{ss'}^{x}-k_{y}\sigma_{ss'}^{y}\right)\nonumber \\
 &  & \qquad\left(d_{xz,s}^{\dagger}(\mbf{k})d_{xz,s'}(\mbf{k})+d_{yz,s}^{\dagger}(\mbf{k})d_{yz,s'}(\mbf{k})\right)\big\rangle\nonumber \\
\Delta_{B_{2u}}^{(4)} & \sim & \big\langle\left(k_{x}\sigma_{ss'}^{y}+k_{y}\sigma_{ss'}^{x}\right)\nonumber \\
 &  & \qquad\left(d_{xz,s}^{\dagger}(\mbf{k})d_{yz,s'}(\mbf{k})-d_{yz,s}^{\dagger}(\mbf{k})d_{xz,s'}(\mbf{k})\right)\big\rangle\nonumber \\
\Delta_{B_{2u}}^{(5)} & \sim & \big\langle\left(k_{x}\sigma_{ss'}^{y}-k_{y}\sigma_{ss'}^{x}\right)\nonumber \\
 &  & \qquad\left(d_{xz,s}^{\dagger}(\mbf{k})d_{yz,s'}(\mbf{k})+d_{yz,s}^{\dagger}(\mbf{k})d_{xz,s'}(\mbf{k})\right)\big\rangle\nonumber \\
\Delta_{B_{2u}}^{(6)} & \sim & \big\langle\left(k_{x}\sigma_{ss'}^{x}+k_{y}\sigma_{ss'}^{y}\right)\nonumber \\
 &  & \qquad\left(d_{xz,s}^{\dagger}(\mbf{k})d_{xz,s'}(\mbf{k})-d_{yz,s}^{\dagger}(\mbf{k})d_{yz,s'}(\mbf{k})\right)\big\rangle\nonumber \\
\label{eq:b2g_a1u_full}
\end{eqnarray}
for $B_{2u}$.

\end{document}